\documentclass[twocolumn]{revtex4-1}
\usepackage{graphicx}
\usepackage{amsmath}
\usepackage[normalem]{ulem}

\draft % marks overfull lines with a black rule on the right

\begin{document}

%\preprint{}

\title{Liquid-liquid phase transitions in dipolar liquids. Insights into Supercooled Water} %Title of paper



\author{Maria Grazia Izzo}
\email[Corresponding author: ]{mariagrazia.izzo@unive.it}
%\thanks{}
\affiliation{Department of Molecular Sciences and Nanosystems, Ca' Foscari University of Venice, Via Torino 155, 30172 Venezia, Italy}


\date{\today}

\begin{abstract}
Dipolar liquids combine two coupled degrees of freedom, translational and dipolar. Intriguingly, these two sectors can reside in distinct thermodynamic states, with the state of one sector possibly favouring a specific state of the other. The onset of ferroelectric order in dipolar liquids is an example of this mutual stabilization. Since this coupling is often encoded in interaction potentials, projecting out one sector to derive an effective interaction for the other can uncover the microscopic mechanism underlying this phenomenon. A coarse-grained theory, solved in the large-dimensional limit, in which the distance between nearest-neighbour particles depends on their dipolar interaction through a density-dependent mechanical compliance shows that a ferroelectric transition can trigger a liquid-liquid transition. It also predicts a first-order liquid-liquid transition within the paraelectric phase that, in turn, can promote ferroelectric ordering. The anisotropic dipolar-dependent soft-core interaction is shown to lead to an effective Jagla-like isotropic two-length-scale interaction, known to support liquid-liquid phase transitions. Here, however, the two-length-scale pattern depends on density and dipolar configuration, and is therefore sensitive to dipolar order. Analysis of supercooled TIP4P/Ice water molecular dynamics simulations shows that first-coordination-shell features of the pair correlation functions are sensitive to dipolar interaction, supporting the foundations of the coarse-grained theory. Comparison of low- and high-density liquid water reveals enhanced spatial anisotropy of the first coordination shell in the low-density liquid, associated, within the theory, with increased effective excluded volume, showing that the liquid-liquid transition involves structural changes within the first shell, beyond reorganization of the interstitial region and tetrahedral order.
\end{abstract}

\maketitle %\maketitle must follow title, authors, abstract and \pacs


\section{Introduction}
\label{Intro}
Dipolar liquids intertwine two degrees of freedom, translational and dipolar. The scenario in which the two degrees of freedom can be in distinct phases is particularly intriguing.
This occurs, for example, in the ferroelectric phase of dipolar liquids \cite{Wei, Levesque, Bartke, Camp, Weis2005}, where dipolar order develops while the translational degrees of freedom remain liquid; in plastic crystals \cite{Lynden, Rescigno}, where translational crystalline order coexists with orientational disorder; and in plastic glasses \cite{Jurkiewicz}, where translational order is preserved while the orientational degrees of freedom become frozen in a disordered state. The coupling between the two sets of degrees of freedom is often encoded in the microscopic interaction potential, as for dipolar liquids. A key question is then whether and how, through this coupling, the state of one set of degrees of freedom can stabilize a particular phase or drive a phase transition in the other. This naturally raises a further question: whether one set of degrees of freedom can be coarse-grained into an effective interaction governing the second set. An example is provided by dipolar liquids, where annealed averaging of the dipolar interaction over the liquid positional degrees of freedom yields an effective ferroelectric-like interaction between the dipoles \cite{Izzo}. The converse issue, namely whether and how the state of the dipolar degrees of freedom can stabilize a particular liquid phase or drive a transition between distinct liquid states of the translational degrees of freedom, has been less explored.
In ferroelectric solids, ferroelectric ordering can induce a distortion of the crystal lattice and, in some systems, a structural phase transition involving a change in the symmetry and shape of the unit cell \cite{Landau,Chaikin,Salje}. In nematic liquid crystals, ferroelectric ordering can similarly couple to distortions of the nematic orientational field \cite{Sebastian}. The counterpart in dipolar liquids would be a liquid–liquid phase transition driven by the onset of macroscopic polarization. This possibility was explored in Ref. \cite{Izzo1} for a dipolar liquid modelling of supercooled water, where a ferroelectric phase transition was shown to promote a transition between two liquid states of different density. The mechanism was rationalized at the macroscopic level through the density dependence of the dipolar contribution to the excess free energy within a classical density functional theory (DFT) scheme. It is therefore of interest to explore the underlying microscopic mechanism and to construct a coarse-grained effective interaction for the translational degrees of freedom arising from their coupling to the dipolar ones. Ref. \cite{Klapp} shows that perfectly polarized dipolar liquids develop pronounced local positional order, with interparticle correlations progressively reflecting the body-centered-tetragonal structure favored by the aligned dipoles. A link between tetrahedral and dipolar ordering has, furthermore, been proposed in supercooled water, where locally tetrahedral structures exhibit short-range dipolar correlations \cite{Yuan}. 
In the absence of long-range translational order and, consequently, of a reference lattice, the liquid-state counterpart of ferroelastic lattice distortion can be sought in the modification of the inherent structure \cite{Stillinger1982,Stillinger1983} of the translational degrees of freedom induced by the dipolar configurations. A local characterization of this modification can be obtained by examining how the dipolar interaction affects the mechanically preferred distance between nearest-neighbor particles. Because this local structural change modifies the effective excluded volume and hence the packing of the liquid, it can translate into a change in a macroscopic thermodynamic quantity such as the density, possibly providing a microscopic mechanism for a liquid–liquid phase transition. The role of directional interactions in stabilizing the local structure of water was investigated through the analysis of its inherent structures \cite{Stillinger}.

A parallel question is whether, already in the isotropic dipolar phase, the dipolar interaction can drive a phase transition in the translational degrees of freedom and whether, in turn, such a transition can promote the onset of ferroelectric order.
Evidence for transitions between distinct fluid states within the dipolar isotropic phase has been reported for dipolar hard spheres at low temperatures and densities \cite{Camp1}. In this regime, dipolar particles associate into head-to-tail chains, and the transition occurs between distinct liquid states differing in density and in the organization of the chain-like structures. At higher densities, chain-like structures can become collectively organized into a ferroelectric liquid state \cite{Levesque}. Moreover, chain formation was found to affect the gas–liquid critical behavior \cite{Bartke}, further highlighting the coupling between dipolar organization and fluid phase behaviour. Whether a liquid-liquid phase transition within the isotropic dipolar phase can promote ferroelectric order, is an open question.
A class of isotropic potentials known to lead to liquid–liquid phase transitions includes two-length-scale core-softened potentials \cite{Stell,Hemmer}, such as the Jagla potential \cite{Jagla,Buldyrev,Ricci}. The Jagla potential consists of a hard core, a finite soft repulsion at short distances, and an attractive interaction at larger distances, thus introducing two characteristic length scales associated with the repulsive core. 
Interestingly, isotropic two-scale core-softened potentials have been proposed as effective interactions \cite{Sadr,Sadr1,Scala} for systems in which anisotropic interactions make the characteristic interparticle distance dependent on molecular orientation, as in water due to directional hydrogen bonding \cite{Stillinger}, suggesting that averaging over molecular orientations of the anisotropic interaction may generate two distinct effective length scales.

The two characteristic length scales of a two-length-scale core-softened potential are expected to leave a fingerprint in the radial pair correlation function, $g(r)$, where $r$ is the radial coordinate \cite{Barraz,Salcedo,Lopez,Partay}. Furthermore, if a liquid-liquid phase transition arises, involving a redistribution between local configurations associated with the two length scales, this redistribution should also be reflected in $g(r)$ \cite{Barraz,Salcedo,Partay}. In particular, when the two length scales represent effective nearest-neighbor contact distances their signature and that of the associated liquid-liquid phase transition are expected to emerge in the onset of the structure associated with the first coordination shell. An increase of population of local configurations associated with the shorter (longer) length scale is expected to shift the onset toward shorter (longer) $r$.
An extensively studied case of a liquid-liquid phase transition, also in connection with two-length-scale core-softened potentials \cite{Sadr,Scala, Ricci, Franzese, Buldyrev}, is the transition between a high-density liquid (HDL) and a low-density liquid (LDL) reported in molecular dynamics (MD) simulations of supercooled water models \cite{Poole,Palmer,Debenedetti,Gartner}. In supercooled water, however, the nearest-neighbor structure of $g(r)$ changes only weakly between LDL and HDL \cite{Foffi,Gallo}. The density difference between the two liquids is instead primarily associated with the increased occupation of interstitial regions between the first and second coordination shells in HDL \cite{Foffi,Gallo}. Furthermore, the increase in density from LDL to HDL is associated with the transformation of the open, unentangled tetrahedral network of LDL, characterized by a larger effective excluded volume and less efficient packing, into the increasingly entangled network characteristic of HDL \cite{Neophytou}. 
A possible explanation for the nearly unchanged location of the peak associated with the first coordination shell across the LDL-HDL transition is that the larger effective excluded volume of LDL reflects a stronger anisotropy of its local structure, which is averaged out in the isotropic radial distribution function.
This picture is consistent with the observation that the intermolecular contact distance in water can depend on directional interactions \cite{Stillinger}.
A minimal description of this effect can be obtained by modeling water as a dipolar liquid, in which the dipolar interaction between a pair of molecules can be attractive or repulsive depending on the relative orientation of their dipoles and on the orientation of their intermolecular vector. Pair configurations corresponding to attractive and repulsive dipolar interactions may thus lead to distinct intermolecular distances, which can be resolved in the radial distribution function conditioned on the sign of the dipolar interaction.
Interestingly, recent MD simulations show that electrostatic interactions render molecular diffusion in water anisotropic in the molecular body frame \cite{Pirnia}.

The paper is organized as follows. Sec. \ref{SecI} introduces a coarse-grained theory of dipolar liquids in which the dipolar interaction renormalizes the characteristic length scale of the short-range repulsive interaction, with a coupling strength set by a density-dependent mechanical compliance.
The theory is formulated in the infinite-dimensional limit, where the virial expansion of the free energy is exactly truncated at second order. Within this framework, the corresponding free-energy functional is derived. Sec. \ref{SecII} analyses the phase behaviour supported by the coarse-grained interaction and establishes a  connection with Jagla-like two-length-scale core-softened potentials. Sec. \ref{SecIII} analyses the radial pair correlations of LDL and HDL states of supercooled water to examine whether the structural signatures predicted by the coarse-grained theory are present. Conclusions are drawn in Sec. \ref{Concl}.

\section{A Coarse-Grained Theory of Dipolar Liquids}
\label{SecI}
\subsection{Adaptive Soft-Core Potential for Dipolar Liquids} \label{SecI1}
Consider a dipolar liquid whose particles interact through a short-range isotropic interaction, $v_0$, whose repulsive core determines an effective excluded volume, together with a dipole-dipole interaction, $w_p$, arising from permanent dipole moments carried by the particles, so that the total potential is 
\begin{equation}
	v=v_0+w_p.
\end{equation}
For purely hard-core interactions, the effective excluded volume coincides with the geometric excluded volume set by the particle diameter, whereas for a generic short-range potential it is an emergent  property of the equilibrium structure of the fluid, reflected in the pair correlation function $g^{(2)}$. The aim of this section is to investigate whether the dipolar interaction can induce a finite renormalization of this effective excluded volume in the thermodynamic limit and, in particular, whether the renormalized excluded volume can be regarded as a functional of the single-particle dipole orientational distribution, thereby becoming sensitive to the onset of ferroelectric order.
In the dipolar liquid, the one-particle density field is
\begin{eqnarray}
	\tilde{\rho}(\mathbf{r},\hat{d})
	=\sum_{i=1}^{N}
	\delta(\hat{d}-\hat{d}_i)
	\delta(\mathbf{r}-\mathbf{r}_i)
	=\rho(\mathbf{r})\zeta(\mathbf{r},\hat{d}),
	\label{rho_tilde}
\end{eqnarray}
where $N$ is the total number of particles, the index $i$ labels the particles,
$\mathbf{r}$ and $\hat{d}$ denote, respectively, the spatial coordinate
and the unit vector associated with dipole orientation,
$\rho(\mathbf{r})$ is the local particle number density marginalized over dipole orientation, and
$\zeta(\mathbf{r},\hat{d})$ is the dipole orientational distribution
function at $\mathbf{r}$.  As appropriate for a liquid, spatial homogeneity is assumed, making $\tilde{\rho}(\mathbf{r},\hat{d})$ independent of $\mathbf{r}$. However, no restriction is imposed on the dipole orientational distribution, thereby allowing for possible dipolar ordering in the liquid. The one-particle density field then factorizes as
\begin{eqnarray}
	\tilde{\rho}
	=\rho\zeta(\hat{d}),
	\label{rho_om}
\end{eqnarray}
where $\rho$ is the particle number density. 
The corresponding Helmholtz free-energy functional can be formally decomposed into an ideal and an excess contribution as
\begin{equation}
	F[\tilde{\rho}]
	=
	F_{\rm id}[\tilde{\rho}]
	+
	\mathcal{F}[\tilde{\rho}], \label{F}
\end{equation}
where
\begin{equation}
	F_{\rm id}[\rho,\zeta]
	=
	\frac{N}{\beta}
	\Big[
	\ln\!\left(\rho\Lambda^3\right)-1
	+
	\int d\hat d\,\zeta(\hat d)\ln\zeta(\hat d)
	\Big],
	\label{F_id}
\end{equation}
with $V$ the macroscopic
volume of the system, $\rho=N/V$, $\beta=(k_BT)^{-1}$ the inverse
temperature, $k_B$ the Boltzmann constant, $T$ the absolute temperature,
and $\Lambda$ the thermal de Broglie wavelength.
The first term on the right-hand side of Eq. \eqref{F_id} is the translational ideal-gas contribution,
whereas the second term is the dipole orientational entropy. The excess free-energy functional is given exactly by \cite{Hansen}
\begin{multline}
	\mathcal{F}[\zeta]
	= \frac{N\rho}{2}
	\int_0^1 d\xi
	\int d\mathbf r_{ij}\,
	d\hat d_i\, d\hat d_j\,
	\zeta(\hat d_i)
	\\
	\times
	g^{(2)}_{\xi}
	(\mathbf r_{ij},\hat d_i,\hat d_j)\,
	\zeta(\hat d_j)\,
	v(\mathbf r_{ij},\hat d_i,\hat d_j).
	\label{extraF_gasid}
\end{multline}
where $\mathbf r_{ij}=\mathbf r_i-\mathbf r_j=r\,\hat r_{ij}$ is the
relative position vector between particles $i$ and $j$.
The auxiliary parameter $\xi\in[0,1]$ continuously switches on the interaction potential from the ideal gas ($\xi=0$) to the fully interacting system ($\xi=1$). The corresponding equilibrium pair correlation function associated with the interaction potential $\xi v$ is denoted by $g^{(2)}_{\xi}$. It encodes all contributions arising from interparticle interactions and entering the excess Helmholtz free-energy functional.
Motivated by the discussion above, it is assumed that the response of the translational degrees of freedom of the many-body environment surrounding a particle pair can be incorporated into a state-dependent effective pair interaction. This interaction is parametrized by an excluded-volume length scale that depends on density and on the dipolar configuration. The dipolar interaction is retained explicitly. At finite dimension, the free-energy functional of the resulting effective system thus does not truncate at second virial order, as the explicit dipolar degrees of freedom generate higher-order virial contributions. As shown below, this truncation becomes exact in the large-dimensional limit, where the higher-order contributions to the free-energy functional vanish.
This construction may be viewed as a liquid-state analogue of ferroelastic coupling \cite{Chaikin}. In crystalline solids, the coupling between dipolar order and lattice strain can modify the characteristic interatomic separations and, consequently, the equilibrium lattice structure.
In a liquid, the analogous effect is represented by a renormalization of the excluded-volume length scale entering the effective pair interaction, which in turn modifies the corresponding pair correlation function $g^{(2)}$.

The representation of collective many-body effects through a state-dependent effective pair interaction has precedents in liquid-state theory. In liquid metals, for example, the screening of ionic interactions by conduction electrons has been incorporated into an effective pair interaction, modeled as a hard-sphere potential with a diameter dependent on the conduction-electron density \cite{Ascarelli}. A density-dependent effective pair potential, furthermore, has been introduced for polymer solutions to incorporate many-body effects while reproducing the pair correlation function of the underlying system and, through the compressibility equation, its equation of state \cite{Bolhuis,Henderson,Chayes}. Although relevant in the present context, the latter approach and the one considered here differ in two respects: (i) the effective potential of Ref. \cite{Bolhuis} is constructed to reproduce the pair correlation functions, while the present coarse-grained potential is assumed to yield the same free energy as the original system; (ii) in Ref. \cite{Bolhuis}, the potential is obtained from pair correlation functions measured in numerical simulations through the hypernetted-chain approximation, while in the present approach its form is motivated by a coarse-grained microscopic picture. The former provides a more realistic representation of the underlying system, while the latter makes explicit the microscopic mechanism that may drive the phase transitions considered below.

To illustrate how dipolar interactions can modify the effective excluded volume, consider an isotropic pair potential $v_0(r)$ with a stable minimum at $r=l_0$, defined by
\begin{equation}
	v_0'(l_0)=0,
	\qquad
	v_0''(l_0)=k_2>0,
\end{equation}
where the prime denotes differentiation with respect to $r$. Assume that each particle carries a permanent dipole moment of magnitude $p$. The corresponding  dipole-dipole interaction in three-dimensions is
\begin{equation}
	w_p^{(3)}(r,\hat r_{ij},\hat d_i,\hat d_j)
	=
	-\frac{p^2}{r^{3}}
	\Delta_{ij},
	\label{w_p0}
\end{equation}
where
\begin{equation}
	\Delta_{ij}
	=
	3(\hat d_i\!\cdot\!\hat r_{ij})
	(\hat d_j\!\cdot\!\hat r_{ij})
	-
	\hat d_i\!\cdot\!\hat d_j. \label{Deltaij}
\end{equation}
The equilibrium zero-temperature particle pair distance is obtained by minimizing the total pair potential
\begin{equation}
	v(r)
	=
	v_0(r)
	-
	\frac{p^2}{r^3}\Delta_{ij}.
\end{equation}
Both the isotropic potential $v_0$ and the dipolar interaction $w_p$ are anharmonic. 
Introducing the displacement $l_{ij}$ and expanding $v_0(r)$ around $l_0$ up to the leading anharmonic term gives
\begin{equation}
	v_0(l_0+l_{ij})
	=
	v_0(l_0)
	+\frac{k_2}{2}l_{ij}^2
	+\frac{k_3}{3!}l_{ij}^3
	+O(l_{ij}^4),
\end{equation}
where $k_3=v_0'''(l_0)$. Solving perturbatively the
stationarity condition for the total pair potential yields
\begin{equation}
	l_{ij}^{\rm eq}
	=
	l_0
	\left[
	-\chi_0\Delta_{ij}
	+c_2\Delta_{ij}^2
	+O(\Delta_{ij}^3)
	\right],
	\label{lij_anh}
\end{equation}
where
\begin{equation}
	\chi_0
	=
	\frac{3p^2}{k_2l_0^5},
	\qquad
	c_2
	=
	-\frac{k_3l_0}{2k_2}\chi_0^2
	-
	\frac{12p^2}{k_2l_0^5}\chi_0.
	\label{chi0_c2}
\end{equation}
For typical soft-core interactions, such as the Lennard-Jones potential,
the isotropic interaction is substantially steeper on the repulsive side
than on the attractive one, implying $k_3<0$. Furthermore, for the Lennard-Jones potential $c_2$ is also positive and the equilibrium distance becomes
a convex function of $\Delta_{ij}$. Physically, this reflects the
asymmetric mechanical response of $v_0$: increasing
the equilibrium pair particle separation is energetically less demanding than reducing
it below $l_0$, where the rapidly increasing short-range
repulsion progressively suppresses further compression. 
Note that $\partial v/\partial \Delta_{ij} = -p^2/r^3 < 0$ is constant and nonzero with respect to $\Delta_{ij}$: $v$ is linear in $\Delta_{ij}$, so no stationary value exists, unlike the interparticle distance. Eq. \eqref{lij_anh} can nonetheless be inverted, giving the $\Delta_{ij}$ compatible with a separation imposed externally, e.g. by a force compressing or decompressing the system. This is a constraint between the two variables, not a minimization over $\Delta_{ij}$: the latter responds to externally driven changes in pair separation, not to self-minimization of the pair energy. 
In a dense liquid, the displacement
of a particle pair towards the equilibrium separation determined by the total
pair potential $v_0+w_p$ generally requires collective rearrangements of
the surrounding particles. Such rearrangements become increasingly difficult with increasing density, reflecting an increase in local stiffness and a corresponding reduction in mechanical compliance. 
The linear structural compliance is therefore generalized as
\begin{equation}
	\chi_0(\rho)
	=
	\frac{3p^2}
	{k_2^{\rm eff}(\rho)l_0^5},
	\label{chi0_rho}
\end{equation}
where $k_2^{\rm eff}(\rho)$ denotes the effective local stiffness.
The structural compliance is positive as a consequence of the local stability of the reference system. For the bare interaction $v_0$, local stability requires $k_2=v_0''(l_0)>0$. For the corresponding coarse-grained potential, the local stability condition entails $k_2^{\rm eff}(\rho)>0$, which, through Eq. \eqref{chi0_rho}, implies $\chi_0(\rho)>0$.
As discussed above, it is physically grounded to assume that
$
\frac{d k_2^{\rm eff}}{d\rho}>0
$, so that finally
\begin{equation}
	\chi_0(\rho)>0,
	\qquad
	\frac{d\chi_0}{d\rho}<0.
	\label{chi0_rho_negative}
\end{equation}
In the parametrization of Eq. \eqref{chi0_rho},collective many-body effects arising from the surrounding liquid environment are indeed encoded solely through the density dependence of the effective stiffness and, therefore, cannot affect the sign of $\chi_0(\rho)$.
A negative value of $\chi_0(\rho)$ is physically possible, but would require a more general coarse-grained description in which many-body effects renormalize not only the effective local stiffness, but also the coupling between $\lambda_{ij}$ and $\Delta_{ij}$. The sign of this effective structural coupling is not constrained by local stability and could therefore yield $\chi_0(\rho)<0$ while $k_2^{\rm eff}(\rho)>0$. The assumption $d\chi_0/d\rho<0$, on the other hand, corresponds, within the coarse-grained parametrization of Eq. \eqref{chi0_rho}, to the physically motivated scenario of compression-induced stiffening, which represents the generally expected mechanical response of a dense liquid upon compression. However, the sign of $d\chi_0/d\rho$ is not constrained by local stability, which only requires $k_2^{\rm eff}(\rho)>0$. A positive derivative, $d\chi_0/d\rho>0$, is therefore also physically possible and would correspond to compression-induced softening. It is anticipated that the density dependence of the structural compliance
introduced in Eq. \eqref{chi0_rho_negative} is consistent with MD simulations
of supercooled TIP4P/Ice water, as shown in Figs. \ref{fig:conditional_gr_water} and \ref{fig:conditional_distance}. In particular, the $\Delta_{ij}$-dependent shifts of the first-coordination-shell
structure are consistent with a positive structural compliance,
$\chi_0>0$, while their enhancement in the LDL relative to the HDL
supports a decrease of $\chi_0$ upon compression,
$d\chi_0/d\rho<0$. Secs. \ref{SecIIa} and \ref{SecIIb} show how the signs of $\chi_0$ and $d\chi_0/d\rho$ affect the phase transitions predicted by the theory.

Guided by this zero-temperature analysis, and retaining only the linear dependence on $\Delta_{ij}$ in Eq. \eqref{lij_anh} for sake of simplicity, the effective pair interaction entering the Helmholtz free-energy functional in Eq. \eqref{extraF_gasid} is assumed to be parameterized by the excluded-volume length scale
\begin{equation}
	\lambda_{ij}=l_0[1-\chi_0(\rho) \ \Delta_{ij}] \label{lambdaij},
\end{equation}
with $\chi_0(\rho) > 0$.
It is worth stressing that the Lennard–Jones potential is not used here as the reference interaction. The reference potential is left unspecified.
An interesting physical interpretation of the effective soft-core length scale $\lambda_{ij}$ introduced in Eq.\eqref{lambdaij} is that it provides a coarse-grained, pair-level description of the inherent structure \cite{Stillinger1982,Stillinger1983} of the translational degrees of freedom of the dipolar liquid.
Indeed, Eq. \eqref{chi0_rho} shows that $\chi_0(\rho)$  reflects the mechanical stiffness of the dipolar liquid and thereby determines $\lambda_{ij}$ in Eq. \eqref{lambdaij}. The latter therefore acts as a coarse-grained structural coordinate of the mechanically relaxed dipolar liquid.

The coarse-grained theory introduced above retains the dipolar degrees of freedom explicitly and may therefore generate higher-order virial contributions at finite dimension $d$. In the limit $d\to\infty$, these contributions vanish under the conditions specified below, and the virial expansion of the excess free-energy functional truncates exactly at second order.
As shown in Ref. \cite{Izzo}, the same
large-$d$ limit already describes the ferroelectric phase transition
of dipolar liquids and serves as the starting point for its extension to
finite $d$.
In the limit $d \rightarrow \infty$, the pair correlation function becomes
\begin{equation}
	g_{\xi}^{(2)}(\mathbf r_{ij},\hat d_i,\hat d_j)=
	e^{-\beta \xi [v_0(r)+w_p(\mathbf r_{ij},\hat d_i,\hat d_j)]}.
	\label{g_dinf}
\end{equation}
Substituting Eq.\eqref{g_dinf} into Eq.\eqref{extraF_gasid} and performing the integration over $\xi$, one obtains
\begin{equation}
	\mathcal{F}[\zeta]
	=
	-\frac{N\rho}{2\beta}
	\int d\mathbf r_{ij}\,
	d\hat d_i\,d\hat d_j\,
	\zeta(\hat d_i)\,
	f(\mathbf r_{ij},\hat d_i,\hat d_j)\,
	\zeta(\hat d_j),
	\label{extraF_mayer_infinite_d}
\end{equation}
where
\begin{equation}
	f(\mathbf r_{ij},\hat d_i,\hat d_j)
	=
	e^{-\beta\left[v_0(r)+w_p(\mathbf r_{ij},\hat d_i,\hat d_j)\right]}-1 \label{Mayertot}
\end{equation}
is the Mayer function associated with the total pair potential.
In the limit $d\rightarrow\infty$, Eq. \eqref{g_dinf} implies that, if $v_0$ is modeled by the generalized soft-sphere potential
\begin{equation}
	v_0(r,\hat r_{ij},\hat d_i,\hat d_j;\rho)
	=
	u_0
	\left(
	\frac{\lambda_{ij}}{r}
	\right)^{\nu d},
	\qquad \nu>1,
	\label{v0softlambdaij}
\end{equation}
where $u_0>0$ sets the energy scale, the resulting pair correlation
function remains associated with an effective excluded volume of characteristic
length $\lambda_{ij}$, the $d$-dimensional generalization of
Eq. \eqref{lambdaij}.
Following Ref. \cite{Izzo}, the generalized Onsager reaction-field
construction is adopted for the dipolar interaction
$w_p$, since the focus is in the occurrence of a genuine bulk
ferroelectric phase transition. The resulting dipolar interaction in $d$-dimension is given by
\begin{widetext}
\begin{equation}
	w_p(r,\hat r_{ij},\hat d_i,\hat d_j;R_c,\epsilon)
	=
	-p^2
	\Big[
	\Big(\frac{l_0}{r}\Big)^d \Delta_{ij}(\hat r_{ij},\hat d_i, \hat d_j)
	+
	\frac{l_0^df_d(\epsilon)}{R_c^d}
	\hat d_i\!\cdot\!\hat d_j
	\Big]
	\theta(r-l_0)\,
	\theta(R_c-r),
	\label{dippotddim}
\end{equation}
\end{widetext}
where 
\begin{equation}
	\Delta_{ij}(\hat r_{ij},\hat d_i, \hat d_j)=[
	d(\hat d_i\!\cdot\!\hat r_{ij})
	(\hat d_j\!\cdot\!\hat r_{ij})
	-
	\hat d_i\!\cdot\!\hat d_j
	] \label{Deltaijd}
\end{equation}
is the $d$-dimensional generalization of the angular invariant introduced in Eq. \eqref{Deltaij} and $f_d(\epsilon)$ is a bounded and adimensional function of the liquid permittivity $\epsilon$.
The factor $\theta(r-l_0)$ removes the short-distance divergence of the dipolar interaction \cite{Izzo} by setting the perturbative dipolar potential to zero for $r<l_0$. The factor $l_0^d$ only sets the scale of the dipolar interaction \cite{Izzo}. Following Ref. \cite{Izzo}, the bulk behavior is recovered in the limit $R_c\to\infty$. From Eq. \eqref{dippotddim} and the property of the bare dipolar potential  $w_p(r,\hat r_{ij},\hat d_i,\hat d_j)=-p^2 \left(\frac{l_0}{r}\right)^d \Delta_{ij}(\hat r_{ij},\hat d_i,\hat d_j)$, $\langle w_p\rangle_{\hat{r}_{ij}}=0$ , 
where $\langle \ \rangle_{\hat{r}_{ij}}$ denotes the average over $\hat{r}_{ij}$ uniformly distributed over the solid angle, one obtains \cite{Izzo}
\begin{multline}
	\int r^{d-1}dr\, d\hat{r}_{ij}\,
	w_p(r,\hat{r}_{ij},\hat{d}_i,\hat{d}_j;R_c,\epsilon)
	\\
	= -p^2 l_0^d f_d(\epsilon)\,
	\hat{d}_i\!\cdot\!\hat{d}_j\,
	\frac{\Omega_d}{d}
	\Big[1-\Big(\frac{l_0}{R_c}\Big)^d\Big].
	\label{wpint0}
\end{multline}
Given a function $g(w_p;r,\hat r_{ij})$ regular in a neighborhood of $w_p=0$, and the Taylor expansion
\begin{equation}
	g(w_p;r,\hat r_{ij})
	=
	g(0;r,\hat r_{ij})
	+
	g_1(r,\hat r_{ij})\,w_p
	+
	O(w_p^2),
\end{equation}
where $g_1(r,\hat r_{ij})
=
\left.
\frac{\partial g(w_p;r,\hat r_{ij})}{\partial w_p}
\right|_{w_p=0}$, if the zeroth-order term of the expansion is absolutely integrable, the remainder $O(w_p^2)$ is bounded in absolute value by $A_d |w_p^2|$, where $A_d$ is a constant  independent of $r$, $\hat r_{ij}$ and $R_c$, and the linear term in $w_p$ is
either absolutely integrable or can be decomposed as
$ g_1(r,\hat r_{ij})\,w_p = C_g^d\,w_p + h_g(r,\hat r_{ij})\,w_p$,
where $C_g^d$ is independent of all $r$, $\hat r_{ij}$ and $R_c$, whereas
$h_g(r,\hat r_{ij})\,w_p$ is absolutely integrable, the following relationship holds
\begin{widetext}
\begin{multline}
	\lim_{R_c\to\infty}
	\int_0^\infty r^{d-1}\,dr
	\int_{\Omega_d} d\hat{r}_{ij}\,
 g\big(
	w_p(r,\hat{r}_{ij},\hat{d}_i,\hat{d}_j;
	R_c,\epsilon);
	r,\hat r_{ij}
	\big)
	\\
	= -C_g^d\,p^2\,f_d(\epsilon)\,
	\frac{\Omega_d}{d}\,
	\hat{d}_i\!\cdot\!\hat{d}_j
	+\Omega_d
	\int_0^\infty r^{d-1}\,dr\,
	\big\langle
	g\big(
	w_p(r,\hat{r}_{ij},\hat{d}_i,\hat{d}_j);
	r,\hat r_{ij}
	\big)
	\big\rangle_{\hat{r}_{ij}}.
	\label{averagerij_gen}
\end{multline}
\end{widetext}
The bare dipolar potential is understood to retain the short-distance regularization $\theta(r-l_0)$, and therefore vanishes for $r<l_0$, consistently with the construction discussed in Ref. \cite{Izzo}.
Eq. \eqref{averagerij_gen} is a generalization of the relationship introduced in Ref. \cite{Izzo}.
The first term on the right-hand side of
Eq. \eqref{averagerij_gen} vanishes in the limit $d\to\infty$ at fixed $p^2$.
When the large-dimensional scalings of $p^2$ and $\rho$ introduced below are adopted, the dielectric permittivity $\epsilon_d$ is assumed to scale consistently so that the reaction-field contribution remains negligible at leading order in the large-$d$ expansion of the free energy.

For convenience, the excess free-energy functional is decomposed into the contributions arising from $v_0$ and $w_p$,
\begin{equation}
	\mathcal F[\zeta]
	=
	\mathcal F_{v_0}[\zeta]
	+
	\mathcal F_{w_p}[\zeta]. \label{Fexdivvowp}
\end{equation}
By rewriting the Mayer function in Eq. \eqref{Mayertot} as
 \begin{equation}
 		f=f_{v_0}+e^{-\beta v_0}\,f_{w_p},
 \end{equation}
where $f_{v_0}=e^{-\beta v_0}-1$ and $f_{w_p}=
e^{-\beta w_p}-1$ are respectively the Mayer functions associated with $v_0$ and $w_p$, one obtains
\begin{align}
	\mathcal{F}_{v_0}[\zeta]
	&=
	-\frac{N\rho}{2\beta}
	\int d\mathbf r_{ij}\,
	d\hat d_i\,d\hat d_j\,
	\zeta(\hat d_i)\,
	f_{v_0}(\mathbf r_{ij},\hat d_i,\hat d_j)\,
	\zeta(\hat d_j),
	\label{F_v0_infinite_d}
	\\[2mm]
	\mathcal{F}_{w_p}[\zeta]
	&=
	-\frac{N\rho}{2\beta}
	\int d\mathbf r_{ij}\,
	d\hat d_i\,d\hat d_j\,
	\zeta(\hat d_i)\,
	e^{-\beta v_0(\mathbf r_{ij},\hat d_i,\hat d_j)}
	\nonumber\\
	&\qquad \qquad \qquad \times
	f_{w_p}(\mathbf r_{ij},\hat d_i,\hat d_j)\,
	\zeta(\hat d_j),
	\label{F_wp_infinite_d}
\end{align}
where $w_p$ is given in Eq. \eqref{dippotddim} following the generalized Onsager's construction. 

\subsection{Excess Free Energy in Large Dimensions}
\label{SecI2}
The potential $v_0$, even in its generalized form in Eq. \eqref{v0softlambdaij}, is not subject to conditional convergence. In full generality Eq. \eqref{F_v0_infinite_d} can then be rephrased as
\begin{multline}
	\mathcal F_{v_0}[\zeta;\rho]
	=
	-\frac{N\rho\,\Omega_d}{2\beta}
	\int_0^\infty r^{d-1}\,dr
	\int d\hat d_i\,d\hat d_j\,
	\zeta(\hat d_i)\,
	\\
	\times \langle
	f_{v_0}(r,\hat r_{ij},\hat d_i,\hat d_j)
	\rangle_{\hat r_{ij}}
	\zeta(\hat d_j)
	\label{F_v0_large_d}
\end{multline}
Eq. \eqref{averagerij_gen}, can be applied to Eq. \eqref{F_wp_infinite_d} by assuming $
	g(w_p;r,\hat r_{ij})
	=
	e^{-\beta v_0(r,\hat r_{ij},\hat d_i,\hat d_j)}
	f_{w_p}$.
The linear term in the Taylor expansion
of $g$ about $w_p=0$ is asymptotic to $-\beta w_p$ as $r\rightarrow\infty$. Its integral is therefore conditionally convergent, while the
remaining contributions are absolutely integrable. Therefore, one obtains
\begin{multline}
	\mathcal F_{w_p}[\zeta;\rho]
	=
	-\frac{N\rho\,\Omega_d}{2\beta}
	\int_0^\infty r^{d-1}\,dr
	\int d\hat d_i\,d\hat d_j\,
	\zeta(\hat d_i)\,
	\\
	\times \langle e^{-\beta v_0(r,\hat r_{ij},\hat d_i,\hat d_j)}
	f_{w_p}(r,\hat r_{ij},\hat d_i,\hat d_j)
	\rangle_{\hat r_{ij}}
	\zeta(\hat d_j).
	\label{F_wp_large_d}
\end{multline}
Within the replicated liquid theory framework \cite{Parisi},
Eqs. \eqref{F_v0_large_d} and \eqref{F_wp_large_d} describe the
single-replica (liquid) free energy, and the average over
$\hat r_{ij}$ corresponds to an annealed average. The average over $\hat r_{ij}$ is taken uniformly over the solid angle.

Following Ref.~\cite{Izzo}, the dipole moment is assumed to scale with $d$ as
\begin{equation}
	p^2=d\,\bar p^{\,2},
	\label{p_scaling}
\end{equation}
so that, in the limit $d\to\infty$, the dipolar contribution to the excess free energy remains of order $O(d)$ and competes with the orientational entropy loss associated with dipolar ordering. Since, under the scaling in Eq. \eqref{p_scaling}, the dipolar interaction remains non-zero only within a thin boundary layer \cite{Izzo} around $r=l_0$, of thickness $O(l_0\log d/d)$, the radial coordinate is conveniently parametrized as
\begin{equation}
	r
	=
	l_0 \Big(
	1+\frac{\log d+h}{d}
	\Big),
	\qquad
	h=O(1),
	\label{radial_scaling}
\end{equation}
so that
\begin{equation}
	p^2\Big(\frac{l_0}{r}\Big)^d
	\underset{d\to\infty}{\longrightarrow}
	\bar p^{\,2}e^{-h}.
	\label{dipolar_scaling_limit}
\end{equation}
As discussed in Ref. \cite{Izzo}, the combined scaling in
Eqs. \eqref{p_scaling} and \eqref{radial_scaling} leads to a well-defined large-dimensional limit of the excess free energy for $h>0$. The
large-$d$ limit of the dipolar interaction is therefore \cite{Izzo}
\begin{equation}
	\beta w_p
	\underset{d\to\infty}{\longrightarrow}
	-\beta\bar p^{\,2} \Delta_{ij}(\hat r_{ij}, \hat d_i,\hat d_j)
	e^{-h}\theta(h),
	\label{wp_large_d_scaling}
\end{equation}
with the reaction-field contribution vanishing as $d \rightarrow \infty$.
Applying the radial scaling in Eq. \eqref{radial_scaling} to $v_0$ and requiring its contribution to the free energy to remain $O(d)$ as $d\rightarrow\infty$ leads to
\begin{equation}
	u_0=d^\nu\bar u_0,
	\label{u0_scaling}
\end{equation}
with $\bar u_0=O(1)$. 
Moreover, a finite, nonvanishing effect of $\lambda_{ij}$ on the soft-core potential as $d \rightarrow \infty$ requires
\begin{equation}
	\chi_0(\rho)
	=
	\frac{\bar\chi_0(\rho)}{d},
	\qquad
	\lambda_{ij}
	=
	l_0
	\Big(
	1-
	\frac{\bar\chi_0(\rho)\Delta_{ij}}{d}
	\Big).
	\label{chi_lambda_scaling}
\end{equation}
Hence,
\begin{equation}
	\Big(\frac{\lambda_{ij}}{l_0}\Big)^{\nu d}
	\underset{d\to\infty}{\longrightarrow}
	e^{-\nu\bar\chi_0(\rho)\Delta_{ij}},
	\qquad
	u_0\Big(\frac{l_0}{r}\Big)^{\nu d}
	\underset{d\to\infty}{\longrightarrow}
	\bar u_0 e^{-\nu h},
	\label{lambda_scaling_limit}
\end{equation}
leading to
\begin{equation}
	\beta v_0
	\underset{d\to\infty}{\longrightarrow}
	\beta \bar u_0\,
	e^{-\nu h-\nu\bar\chi_0(\rho)\Delta_{ij}}.
	\label{v0_h}
\end{equation}
From Eq. \eqref{radial_scaling} it follows
\begin{equation}
	r^{d-1}\,dr
	\underset{d\to\infty}{\longrightarrow}
	l_0^d e^h\,dh.
	\label{radial_measure_scaling}
\end{equation}
To evaluate $\mathcal{F}_{v_0}$, it is convenient to perform the change of variables from $h$ to $\tilde h$ with $h=\tilde h -\bar \chi_0(\rho) \Delta_{ij}$, which is equivalent to parameterizing  in the limit $d \rightarrow \infty$ the radial coordinate with respect to $\lambda_{ij}$, 
\begin{equation}
	r
	=
	\lambda_{ij}
	\Big(
	1+\frac{\log d+\tilde h}{d}
	\Big),
	\qquad
	\tilde h=O(1). \label{radial_scaling_lambda}
\end{equation}
Eqs. \eqref{v0_h} and \eqref{radial_measure_scaling} become
\begin{equation}
	\beta v_0
	\underset{d\to\infty}{\longrightarrow}
	\beta \bar u_0\,
	e^{-\nu \tilde h};
	\label{v0_htilde}
\end{equation}
\begin{equation}
	r^{d-1}\,dr
	\underset{d\to\infty}{\longrightarrow}
	\lambda_{ij}^d e^{\tilde h}\,d \tilde{h}.
	\label{radial_measure_scaling_tilde}
\end{equation} 
Since in the integral defining $\mathcal F_{v_0}$, $\tilde h$ is a dummy integration variable, the tilde is omitted and the integration variable denoted by $h$.   
The excess free energy associated to $v_0$ in Eq. \eqref{F_v0_large_d} thus takes the form
\begin{multline}
	\mathcal F_{v_0}[\zeta;\rho]
	=
	-\frac{N\rho\, B_2^{\mathrm{HS}}}{\beta}d
	\int_{-\infty}^{\infty}dh\,e^h \ f_{v_0}(h)
	\\
	\times
	\int d\hat d_i\,d\hat d_j\,
	\zeta(\hat d_i) \ 
	\bar\eta(\hat d_i \cdot \hat d_j;\rho)
	\zeta(\hat d_j),
	\label{F_v0_h}
\end{multline}
where $B_2^{\mathrm{HS}}=\Omega_d l_0^d/(2d)$ is the hard-sphere second virial coefficient and the reduced density $\rho B_2^{\mathrm{HS}}$ is kept $O(1)$. The virial expansion is therefore exactly truncated at second order in the limit $d\to\infty$, provided that
$
\int d\hat d_i\,d\hat d_j\,
\zeta(\hat d_i)\,
\bar\eta(\hat d_i,\hat d_j;\rho)\,
\zeta(\hat d_j)
$
remains finite, where
\begin{equation}
	\bar\eta(\hat d_i \cdot \hat d_j;\rho)
	=
	l_0^{-d}
	\lim_{d\to\infty}
	\Big\langle \Big(
	\frac{\lambda_{ij}(\hat r_{ij},\hat d_i,\hat d_j)}{l_0}\Big)^{\,d}
	\Big \rangle_{\hat r_{ij}}.
	\label{etabar}
\end{equation}
From Eq. \eqref{lambda_scaling_limit},
\begin{equation}
	\bar\eta(\hat d_i \cdot \hat d_j;\rho)
	=
	\lim_{d\to\infty}
	\langle e^{-\bar \chi_0(\rho) \Delta_{ij}(\hat r_{ij}, \hat d_i, \hat d_j)}
	\rangle_{\hat r_{ij}}.
	\label{etabar1}
\end{equation} 
Introducing the set of variables 
\begin{equation}
	\mathbf t_{ij}
	=
	(\theta_i,\theta_j)
	=
	(\hat d_i\cdot\hat r_{ij},
	\hat d_j\cdot\hat r_{ij}), \label{tij}
\end{equation}
considering the expression of $\Delta_{ij}$ in Eq. \eqref{Deltaijd},
and following the same steps of Ref. \cite{Izzo}, one obtains
\begin{equation}
	\bar{\eta}(\hat d_i \cdot \hat d_j;\rho)
	=
	\int_{-\infty}^{\infty} d\tilde{\mathbf{t}}_{ij}\,
	\frac{
		e^{-\frac{1}{2}\tilde{\mathbf{t}}_{ij}\mathbf{G}^{-1}\tilde{\mathbf{t}}_{ij}^{T}}
	}
	{2\pi(\det\mathbf{G})^{1/2}}
	e^{-\bar{\chi}_0(\rho)(\tilde{\theta}_i\tilde{\theta}_j-\hat d_i\cdot\hat d_j)},
	\label{etabardinf}
\end{equation}
where
$\tilde{\mathbf t}_{ij}=\sqrt{d}\,\mathbf t_{ij}$,
$\tilde{\mathbf t}_{ij}^{\,T}$ denotes the transpose of
$\tilde{\mathbf t}_{ij}$,
and $\mathbf G$ is the $(2\times2)$ Gram matrix with entries
$G_{mn}=\hat d_m\!\cdot\!\hat d_n$. It follows from Eq. \eqref{etabardinf} that $\bar{\eta}(\hat d_i,\hat d_j;\rho)$, and hence $v_{\rm ex}$, remains finite, $\forall \zeta(\hat d)$, if
\begin{equation}
	0<\bar{\chi}_0(\rho)<\frac{1}{2}. \label{chiintexistFv0}
\end{equation}
This is therefore a necessary condition for the virial expansion of the excess free energy to be exactly truncated at second order in the limit $d\to\infty$.
Following Ref. \cite{Izzo}, one observes that the integral in Eq. \eqref{etabardinf} coincides with the moment-generating function $M_{\tilde Z}(x)=\langle
e^{x\tilde Z} \rangle$ of the random variable
\begin{equation}
	\tilde Z
	=
	\tilde\theta_i\tilde\theta_j
	-
	\hat d_i\cdot\hat d_j,
	\label{Ztilde}
\end{equation}
evaluated with respect to the bivariate Gaussian distribution of $\tilde{\mathbf t}_{ij}$. Its explicit expression is \cite{Izzo}
\begin{equation}
	M_{\tilde Z}(x)
	=
	e^{-x\hat d_i\cdot\hat d_j
		-\frac{1}{2}\log\!\left[
		1-2x\,\hat d_i\cdot\hat d_j
		-x^2\left(1-(\hat d_i\cdot\hat d_j)^2\right)
		\right]}.
	\label{MGF}
\end{equation}
Here, $\langle \ \rangle$ denotes the expectation with respect to the probability distribution of the random variable $\tilde Z$, whose moment-generating function is given in Eq. \eqref{MGF}. Therefore,
\begin{equation}
	\bar{\eta}(\hat d_i\cdot\hat d_j;\rho)
	=
	M_{\tilde Z}(-\bar\chi_0(\rho)).
	\label{etabardinfMz}
\end{equation}    
The excess free energy associated with $w_p$ is obtained from Eq. \eqref{F_wp_large_d} by using the radial measure in Eq. \eqref{radial_measure_scaling}, yielding \cite{Izzo}
\begin{multline}
	\mathcal F_{w_p}[\zeta;\rho]
	=
	-\frac{N\rho\, B_2^{\mathrm{HS}}}{\beta}d
	\int_{-\infty}^{\infty}dh\,e^h
	\\
	\times
	\int d\hat d_i\,d\hat d_j\,
	\zeta(\hat d_i)\,
	\bar\phi(h,\hat d_i \cdot \hat d_j;\rho)\,
	\zeta(\hat d_j),
	\label{F_wp_h}
\end{multline}
where
\begin{multline}
	\bar\phi(h,\hat d_i \cdot \hat d_j;\rho)
	=\\
	\lim_{d\to\infty}
	\langle
	e^{-\beta v_0(h,\hat r_{ij},\hat d_i,\hat d_j)}
	[
	e^{-\beta w_p(h,\hat r_{ij},\hat d_i,\hat d_j)}
	-1
	]
	\rangle_{\hat r_{ij}}.
	\label{phibar}
\end{multline}
Using Eqs. \eqref{wp_large_d_scaling} and \eqref{v0_h}, one finds
\begin{widetext}
\begin{equation}
	\bar{\phi}(h, \hat d_i \cdot \hat d_j;\rho)
	=
	\int_{-\infty}^{\infty} d\tilde{\mathbf{t}}_{ij}\,
	\frac{
		e^{-\frac{1}{2}\tilde{\mathbf{t}}_{ij}\mathbf{G}^{-1}\tilde{\mathbf{t}}_{ij}^{T}}
	}
	{2\pi(\det\mathbf{G})^{1/2}}
	e^{-\beta \bar u_0\,
		e^{-\nu h-\nu\bar\chi_0(\rho)(\tilde{\theta}_i\tilde{\theta}_j-\hat d_i\cdot\hat d_j)}} \Big(e^{\beta\bar p^{\,2} (\tilde{\theta}_i\tilde{\theta}_j-\hat d_i\cdot\hat d_j)
		e^{-h}\theta(h)} -1\Big).
	\label{phibardinf}
\end{equation}
\end{widetext}
The convergence of the integral in Eq. \eqref{phibardinf}, and hence the exact truncation of the virial expansion at second order in the limit $d\to\infty$, requires
\begin{equation}
	\beta\bar p^{\,2}<\frac{1}{2}. \label{condp}
\end{equation}
Eq. \eqref{condp} gives a necessary condition for the virial expansion to truncate exactly at second order in the limit $d \rightarrow \infty$ \cite{Izzo}, in addition to the bound on $\bar \chi_0(\rho)$ in Eq. \eqref{chiintexistFv0}.  
Expanding the first exponential in Eq. \eqref{phibardinf}, the soft-core Boltzmann factor, in powers of $\bar\chi_0(\rho)$ about $\bar\chi_0(\rho)=0$, and using the random variable $\widetilde Z_{ij}$ in Eq. \eqref{Ztilde}, yields
\begin{multline}
	e^{-\beta\bar u_0
		e^{-\nu h-\nu\bar\chi_0(\rho)\widetilde Z_{ij}}}
	=
	\sum_{n=0}^{\infty}(-1)^n
	\frac{(\nu\bar\chi_0(\rho))^n}{n!}
	\widetilde Z_{ij}^{\,n}
	\\
	\times
	\Big.
	\frac{d^n}{dx^n}
	e^{-\beta\bar u_0e^{-\nu h+x}}
	\Big|_{x=0}.
	\label{chi_series}
\end{multline}
Combining Eqs. \eqref{chi_series} and \eqref{phibardinf} gives the formal expansion
\begin{multline}
	\bar{\phi}(h,\hat d_i\!\cdot\!\hat d_j;\rho)
	=
	\sum_{n=0}^{\infty}(-1)^n
	\frac{(\nu\bar\chi_0(\rho))^n}{n!}
	\Big.
	\frac{d^n}{dx^n}
	e^{-\beta\bar u_0e^{-\nu h+x}}
	\Big|_{x=0}
	\\
	\times
	\Big[
	M_{\tilde Z}^{(n)}
	\!\left(
	\beta\bar p^{\,2}e^{-h}\theta(h)
	\right)
	-
	M_{\tilde Z}^{(n)}(0)
	\Big].
	\label{phibar_chi_series}
\end{multline}
where the $n$th derivative of the moment-generating function
$M_{\tilde Z}(s)$ of the random variable $\widetilde Z$, defined in
Eq. \eqref{Ztilde}, is
\begin{equation}
	M_{\tilde Z}^{(n)}(x)
	=
	\left.
	\frac{d^nM_{\tilde Z}(s)}{ds^n}
	\right|_{s=x}
	=\langle
	\tilde Z^{\,n}
	e^{x\tilde Z}
	\rangle.
\end{equation}
It is convenient to introduce the polynomials $\mathcal P_n(y)$ through
\begin{equation}
	\frac{d^n}{dx^n}
	e^{-y}
	=
	e^{-y}
	\mathcal P_n(y); \qquad y=\beta\bar u_0e^{-\nu h+x},\label{y}
\end{equation}
so that
\begin{equation}
	\frac{d}{dx}
	=
	y\frac{d}{dy}.
\end{equation}
Differentiating Eq. \eqref{y}, one obtains
\begin{equation}
	\frac{d^{n+1}}{dx^{n+1}}
	e^{-\beta\bar u_0e^{-\nu h+x}}
	=
	y\frac{d}{dy}
	\left[
	e^{-y}\mathcal P_n(y)
	\right],
\end{equation}
from which the recursion relation follows,
\begin{equation}
		\mathcal P_0(y)=1,
		\qquad
		\mathcal P_{n+1}(y)
		=
		y\left[
		\mathcal P_n'(y)-\mathcal P_n(y)
		\right].
	\label{Pn_recursion}
\end{equation}
Eq. \eqref{phibar_chi_series} becomes
\begin{widetext}
\begin{equation}
	\bar{\phi}(h,\hat d_i\!\cdot\!\hat d_j;\rho)
	=
	e^{-\beta\bar u_0e^{-\nu h}}
	\sum_{n=0}^{\infty}(-1)^n
	\frac{(\nu\bar\chi_0(\rho))^n}{n!}
	\mathcal P_n\!\left(
	\beta\bar u_0e^{-\nu h}
	\right)
	\left[
	M_{\tilde Z}^{(n)}
	\!\left(
	\beta\bar p^{\,2}e^{-h}\theta(h)
	\right)
	-
	M_{\tilde Z}^{(n)}(0)
	\right].
	\label{phibar_chi_series1}
\end{equation}
\end{widetext}
Notice that both $\mathcal F_{v_0}$ and $\mathcal F_{w_p}$ involve only pairwise scalar products of  $\hat d_i$ and $\hat d_j$.

\subsection{Effective Excluded Volume and Equilibrium Density} \label{SecIC}
Since \cite{Izzo}
$\langle\Delta_{ij}\rangle_{\hat r_{ij}}=0$, Jensen's inequality implies
\begin{equation}
	\bar\eta_{\chi_0}(\hat d_i\!\cdot\!\hat d_j;\rho)
	\ge
	\bar\eta_0(\hat d_i\!\cdot\!\hat d_j).
	\label{etadis}
\end{equation}
Being
$f_{v_0}(h)\le0$, the weight
$-e^hf_{v_0}(h)$ is non-negative such as 
$\zeta(\hat d_i)\zeta(\hat d_j)$, it follows that
\begin{equation}
	\mathcal F_{v_0}^{\bar \chi_0}[\zeta;\rho]
	-
	\mathcal F_{v_0}^0[\zeta;\rho]
	\ge0.
	\label{Fv0dis}
\end{equation}
Hence, the pair contact deformation induced by $w_p$
carries a positive free-energy cost in
$\mathcal F_{v_0}$.
Moreover, from Eqs. \eqref{v0_h}-\eqref{etabar1} and Eq. \eqref{etabardinfMz} follows that \begin{equation} 
	\frac{\partial^2\mathcal F_{v_0}} {\partial\bar\chi_0^2} \ge0. \end{equation} 
Indeed, 
\begin{equation} \frac{\partial^2\bar\eta} {\partial\bar\chi_0^2} = \langle \tilde Z^{\,2} e^{-\bar\chi_0\tilde Z} \rangle \ge0. \label{etader2} 
\end{equation} 
and $-e^hf_{v_0}(h)$ is non-negative.

The effective excluded volume of the dipolar liquid is defined as
\begin{equation}
	 V_{\rm ex}[\zeta;\rho]=\frac{\Omega_d}{d}\int d\hat d_i\,d\hat d_j\,
	\zeta(\hat d_i)\, \langle \lambda_{ij}^d(\hat d_i, \hat d_j;\rho) \rangle_{\hat r_{ij}} \zeta(\hat d_j), \label{Vex_def}
\end{equation}
following Eq. \eqref{lambdaij}. In the large-dimensional limit, it becomes
\begin{equation}
	\bar V_{\rm ex}[\zeta; \rho]=\frac{\Omega_d}{d} l_0^d
	\int d\hat d_i\,d\hat d_j\,
	\zeta(\hat d_i)\,
	\bar\eta(\hat d_i,\hat d_j;\rho)\,
	\zeta(\hat d_j), \label{Vex_op}
\end{equation}
see Eqs. \eqref{etabar} and \eqref{etabar1}.
From Eq. \eqref{etadis}, any nonzero value of $\bar\chi$ increases $\bar V_{\rm ex}$. 
Furthermore, following Eqs. \eqref{etader2} and \eqref{etadis},
$\bar\eta$ is convex and has its minimum at $\bar\chi_0=0$.
Consequently, for $\bar\chi_0>0$, $\frac{\partial\bar\eta}{\partial\bar\chi_0}\geq0$.
Since $\zeta(\hat d)$ entering Eq.~\eqref{Vex_op} is non-negative,
\begin{equation}
	\frac{\partial\bar V_{\rm ex}}
	{\partial\bar\chi_0}
	\geq0,
	\qquad \bar\chi_0>0.
	\label{Vex_chi_monotonic}
\end{equation}
Thus, decreasing the density increases $\bar\chi_0$, according to
Eq. \eqref{chi0_rho_negative}, and therefore increases the effective
excluded volume.
By defining an effective hard-sphere second virial
coefficient associated with $v_0$,
\begin{equation}
B_2^{\mathrm{v_0}}[\zeta;\rho]= \frac{\bar V_{ex}[\zeta;\rho]}{2},
	\label{B2v0}
\end{equation}
Eq. \eqref{F_v0_h} can be rephrased as
\begin{equation}
	\mathcal F_{v_0}[\zeta;\rho]
	=
	-\frac{N\rho\, B_2^{\mathrm{v_0}}[\zeta;\rho]}{\beta}d
	\int_{-\infty}^{\infty}dh\,e^h \ f_{v_0}(h).
	\label{F_v0_hB2v0}
\end{equation}
Thus, the increase of $\mathcal F_{v_0}$ established in
Eq. \eqref{Fv0dis} follows directly from the increase of the effective
excluded volume. 
As a useful property, in the following it is proved that the first derivative of $\mathcal F_{w_p}$ in $\bar \chi_0$ is 
\begin{equation}
	\frac{\partial \mathcal F_{w_p}}
	{\partial\bar\chi_0}<0. \label{derFwpchi0}
\end{equation}
It is straightforward to obtain
\begin{widetext}
\begin{multline}
	\frac{\partial \mathcal F_{w_p}}
	{\partial\bar\chi_0}
	=-
	N\rho B_2^{\mathrm{HS}}d\,\nu\bar u_0
	\int_{-\infty}^{\infty}dh\,e^{(1-\nu)h}\int d\hat d_i\,d\hat d_j\zeta(\hat d_i)
	 \zeta(\hat d_j)\langle
	\tilde Z\,e^{-\nu\bar\chi\tilde Z}
	e^{-\beta\bar u_0e^{-\nu h-\nu\bar\chi_0\tilde Z}}
	(
	e^{\beta\bar p^{\,2}e^{-h}\theta(h)\tilde Z}-1)
	\rangle.
	\label{dFwp_dchi_general}
\end{multline}
\end{widetext}
For $h<0$ the integrand vanishes, whereas for $h>0$, $	\tilde Z
(
e^{\beta\bar p^{\,2}e^{-h}\widetilde Z}-1
)
\ge0$, $\forall \widetilde Z$, while all remaining factors are strictly
positive. Therefore, Eq. \eqref{derFwpchi0} follows 
for $\bar p^{\,2}>0$ and a nondegenerate distribution of
$\widetilde Z$. 
Thus, increasing $\bar \chi_0$ reduces the contribution of $\mathcal F_{w_p}$ to the total free energy, independently of the
orientational distribution $\zeta(\hat d)$,  thereby providing a free-energy gain. Because $\chi_0(\rho)$ decreases with increasing density, as in Eq. \eqref{chi0_rho_negative} this gain is
progressively weakened with increasing density.
The opposite monotonic behavior of $\mathcal F_{v_0}$ and
$\mathcal F_{w_p}$ with respect to $\bar \chi_0$ shows that the
structural response is governed by a genuine energetic competition.
While increasing $\bar \chi_0$ raises the reference free energy,
it simultaneously lowers the dipolar contribution. This energetic competition supports the interpretation of $\chi_0$ as a microscopic measure of the local mechanical compliance.

The equilibrium density is determined in the isobaric ensemble.
At fixed T and pressure, P, the equilibrium state is obtained
by minimizing the Gibbs free energy per particle
\begin{equation}
	g[\zeta;\rho]
	=
	\frac{F[\zeta;\rho]}{N}
	+
	\frac{P}{\rho},
	\label{g_isobaric}
\end{equation}
with respect to the orientational distribution $\zeta(\hat d)$ and the
particle density $\rho$.
In the limit $d\to\infty$,
\begin{equation}
	\frac{\beta F[\zeta;\rho]}{N}
	=
	\log(\rho\Lambda^d)-1
	+
	\int d\hat d\,
	\zeta(\hat d)\ln\zeta(\hat d)
	+
	\rho\,
	\mathcal B_2
	\left[
	\zeta;\bar\chi_0(\rho)
	\right],
	\label{F_isobaric_large_d}
\end{equation}
where
\begin{equation}
	\mathcal B_2
	\left[
	\zeta;\bar\chi_0(\rho)
	\right]
	=
	\frac{\beta\mathcal F[\zeta;\rho]}
	{N\rho}.
\end{equation}
The equilibrium conditions are
\begin{equation}
	\frac{\delta g}{\delta\zeta(\hat d)}=0,
	\qquad
	\frac{\partial g}{\partial\rho}=0,
	\label{isobaric_stationarity}
\end{equation}
together with $\int d\hat d\,\zeta(\hat d)=1$. The stationarity condition with respect to $\rho$ yields
\begin{equation}
	\frac{1}{\rho}
	+
	\mathcal B_2
	+
	\rho
	\frac{\partial\mathcal B_2}
	{\partial\bar\chi_0}
	\frac{d\bar\chi_0}{d\rho}
	-
	\frac{\beta P}{\rho^2}
	=
	0.
	\label{rho_stationarity}
\end{equation}
Equivalently,
\begin{equation}
	\beta P
	=
	\rho_{eq}
	+
	\rho_{eq}^2
	\mathcal B_2
	+
	\rho_{eq}^3
	\frac{\partial\mathcal B_2}
	{\partial\bar\chi_0}
	\frac{d\bar\chi_0}{d\rho}.
	\label{EOS_chi_rho}
\end{equation}

\section{Ferroelectric and liquid-liquid phase transitions in dipolar liquids} \label{SecII}
\subsection{From ferroelectric to liquid-liquid phase transition} \label{SecIIa}
In the following, it is investigated whether the Helmholtz free energy, given by
Eqs. \eqref{F}, \eqref{F_id}, \eqref{F_v0_h}, \eqref{Fexdivvowp} and \eqref{F_wp_h},
undergoes a ferroelectric phase transition. To this end, it is analyzed which value of $\hat d_i \cdot \hat d_j$ minimize
$\mathcal{F}_{v_0}$, Eq. \eqref{F_v0_h}, and
$\mathcal{F}_{w_p}$, Eq. \eqref{F_wp_h}. The onset of ferroelectric order is determined by the balance in the free energy between a possible reduction of the excess free energy due to dipole alignment and the increase associated with the loss of dipole rotational entropy.

The dependence on $\zeta (\hat d)$ of $\mathcal F_{v_0}$ is entirely
encoded in the effective excluded-volume factor in Eq. \eqref{etabardinfMz}.
From Eq. \eqref{MGF} then
\begin{equation}
	\bar\eta(\hat d_i\!\cdot\!\hat d_j;\rho)
	=
	\frac{e^{\bar \chi_0 \hat d_i\!\cdot\!\hat d_j}}
	{\sqrt{1+2\bar \chi_0 d_i\!\cdot\!\hat d_j-\bar \chi_0^2(1-{\hat d_i\!\cdot\!\hat d_j}^2)}}.
	\label{eta_q}
\end{equation}
Differentiating Eq. \eqref{eta_q} with respect to $\hat d_i\!\cdot\!\hat d_j$ gives
\begin{equation}
	\frac{\partial}{\partial \ \hat d_i\!\cdot\!\hat d_j}
	\ln\bar\eta(\hat d_i\!\cdot\!\hat d_j;\rho)
	=
	\frac{\bar \chi_0^2
		[
		\hat d_i\!\cdot\!\hat d_j-\bar \chi_0(1-{\hat d_i\!\cdot\!\hat d_j}^2)
		]}
	{1+2\bar \chi_0 \hat d_i\!\cdot\!\hat d_j-\bar\chi_0^2(1-{\hat d_i\!\cdot\!\hat d_j}^2)}.
	\label{dlogeta_dq}
\end{equation}
Since the denominator is strictly positive for
$0<\bar\chi_0<1/2$, the stationary points satisfy
\begin{equation}
	\hat d_i\!\cdot\!\hat d_j-\bar\chi_0(1-{\hat d_i\!\cdot\!\hat d_j}^2)=0.
\end{equation}
The only solution in the physical interval $-1\le \hat d_i\!\cdot\!\hat d_j\le1$ is
\begin{equation}
	{\hat d_i\!\cdot\!\hat d_j}^*
	=
	\frac{\sqrt{1+4\bar\chi_0^2}-1}{2\bar\chi_0}, \label{d_id_jetastaz}
\end{equation}
which is positive for $\bar\chi_0>0$. It corresponds to a minimum of
$\bar\eta$.
Hence, $\bar\eta(\hat d_i\!\cdot\!\hat d_j;\rho)$ possesses a unique global minimum at
$\hat d_i\!\cdot\!\hat d_j={\hat d_i\!\cdot\!\hat d_j}^*>0$. This minimum is attainable by a suitable orientational distribution. From Eq. \eqref{F_v0_h}, it follows that $\mathcal F_{v_0}$ is minimized in a dipole phase where $\hat d_i \cdot \hat d_j>0$. Since $ij$ denotes an arbitrary pair of dipoles, a state characterized by $\hat d_i \cdot \hat d_j >0$ has
nonvanishing macroscopic polarization, i.e. it is a ferroelectric state.
For $\chi_0=0$, $\mathcal F_{w_p}$ is globally minimized by the
ferroelectric state, with complete dipolar alignment,
$\hat d_i\!\cdot\!\hat d_j=1$, as demonstrated in Ref. \cite{Izzo}. At finite $\chi_0$, however,
Eq. \eqref{phibardinf} shows that the dependence of $\bar\phi(h,\hat d_i \cdot \hat d_j;\rho)$ on
$\hat d_i\!\cdot\!\hat d_j$ is no longer determined solely by
$M_{\widetilde Z}
\!\left(
\beta\bar p^{\,2}e^{-h}\theta(h)
\right)$, as for $\bar \chi_0=0$ \cite{Izzo}.
A general extension of the $\bar \chi_0=0$ result is therefore nontrivial.
To investigate whether the ferroelectric minimum is
preserved for weak $\bar \chi_0$, consider the first-order term, $\bar \phi^{(1)}$,
of the expansion in $\bar \chi_0$ around $\bar \chi_0=0$ of Eq. \eqref{phibar_chi_series}. 
Its
derivative with respect to $\hat d_i\!\cdot\!\hat d_j$ at $\hat d_i\!\cdot\!\hat d_j=0$ is
\begin{multline}
	\left.
	\frac{\partial\bar\phi^{(1)}}
	{\partial(\hat d_i\!\cdot\!\hat d_j)}
	\right|_{\hat d_i\cdot\hat d_j=0}
	=
	\nu\bar\chi_0(\rho)\,
	\beta\bar u_0e^{-\nu h}
	\\
	\times
	e^{-\beta\bar u_0e^{-\nu h}}
	\left.\frac{3x^2}
	{(1-x^2)^{5/2}}
	\right|_{x=\beta\bar p^{\,2}e^{-h}\theta(h)}>0.
	\label{dphibar_first_order}
\end{multline}
Thus, $\bar\phi^{(1)}$ increases upon moving from $\hat d_i\!\cdot\!\hat d_j=0$ toward positive values. Since $\bar\phi^{(1)}$ enters the first-order contribution to $\mathcal F_{w_p}$ with a negative sign, this corresponds to a decrease of the dipolar contribution to the free energy.
Therefore, for sufficiently small values of $\chi_0$, the first-order correction also makes $\mathcal{F}_{w_p}$ decrease upon the development of a small nonzero macroscopic polarization from the paraelectric state.
The total excess free energy $\mathcal F_{v_0}+\mathcal F_{w_p}$ in this regime thus favours dipole ordering. This decrease in the excess free energy can overcome the competing increase in the orientational-entropy contribution, allowing the onset of a ferroelectric phase transition. Consider now the case of small negative $\chi_0$. In this case, the right-hand side of Eq. \eqref{dphibar_first_order} changes sign, so that $\bar\phi^{(1)}$ decreases upon moving from
$\hat d_i\!\cdot\!\hat d_j=0$ toward positive values. As discussed
above, this instead favors negative values of
$\hat d_i\!\cdot\!\hat d_j$. Likewise, the stationary point of
$\bar\eta$ in Eq. \eqref{d_id_jetastaz} changes sign with $\chi_0$ and is therefore negative for $\chi_0<0$.
Thus both $\mathcal F_{v_0}$ and $\mathcal F_{w_p}$
favor negative values of $\hat d_i\!\cdot\!\hat d_j$ for sufficiently small negative
$\chi_0$, possibly supporting an antiferroelectric phase
transition. Thus, the sign of $\chi_0$ qualitatively affects the ferroelectric phase transition.

A quantitative description of the ferroelectric phase transition is obtained by restricting $\zeta(\hat d)$ to the family
\begin{equation}
	\zeta(\hat d)
	=
	\frac{1}{Z_d(\boldsymbol{\delta})}
	e^{d\boldsymbol{\delta}\cdot\hat d}; \qquad 	Z_d(\boldsymbol{\delta})
	=
	\int d\hat d\,
	e^{d\boldsymbol{\delta}\cdot\hat d},
	\label{zeta}
\end{equation}
with $\mathbf \delta=\delta \hat \delta$ and $0 \leq \delta \leq 1$.
As shown in Ref. \cite{Izzo}, this ansatz describes the paraelectric and ferroelectric phases in the limit $d\rightarrow\infty$, corresponding respectively to $\boldsymbol{\delta}=0$ and $\boldsymbol{\delta}\neq0$. Following Ref. \cite{Izzo}, the polarization per particle is a function of $\mathbf \delta$, i.e.
\begin{equation}
	\boldsymbol{\bar p}
	=
	\bar u(\delta)\,\hat\delta; \qquad 	\bar u(\delta)
	=
	\frac{\sqrt{1+4\delta^2}-1}{2\delta}.
\end{equation}
For $\delta\ll1$,
\begin{equation}
	\boldsymbol{\bar p}
	=
	\boldsymbol{\delta}
	+
	O(\delta^3),
\end{equation}
so that $\boldsymbol{\delta}$ is proportional to the macroscopic
polarization and therefore serves as the ferroelectric order parameter.
Within the ansatz \eqref{zeta}, $F_{\rm id}$ is \cite{Izzo}
\begin{equation}
	\frac{\beta F_{\rm id}(\delta,\rho)}{N}
	=
	\log(\rho\Lambda^d)-1
	-\log\Omega_d
	-\frac{d}{2}\log(1-\bar u^2)
	+
	o(d).
	\label{Fid_ubar}
\end{equation}
For $\delta \ll 1$,
\begin{equation}
	\frac{\beta F_{\rm id}}{N}
	=
	\frac{\beta F_{\rm id}(0)}{N}
	+
	d\Big[
	\frac{1}{2}\delta^2
	-\frac{3}{4}\delta^4
	+
	O(\delta^6)
	\Big].
	\label{Fid_delta_expansion}
\end{equation}
Furthermore, Laplace's saddle-point method \cite{Izzo} yields
\begin{multline}
	\mathcal F_{v_0}(\rho,\delta)
	=
	-\frac{N\rho B_2^{\mathrm{HS}}}{\beta}\,d
	\int_{-\infty}^{\infty}dh\,e^h f_{v_0}(h)\,
	\\
	\times M_{\widetilde Z}
	\left(
	-\bar\chi_0(\rho)
	\right)
	\bigg|_{\hat d_i\cdot\hat d_j=\bar u^2(\delta)},
	\label{Fv0_delta}
\end{multline}
with $M_{\tilde Z}(x)$ given in Eq. \eqref{MGF}. Analogously,
\begin{equation}
	\mathcal F_{w_p}(\rho,\delta)
	=
	-\frac{N\rho B_2^{\mathrm{HS}}}{\beta}\,d
	\int_{-\infty}^{\infty}dh\,e^h\,
	\bar\phi
	\!\left(
	h,\bar u^2(\delta);\rho
	\right).
	\label{Fwp_delta}
\end{equation}
Expanding Eq. \eqref{Fv0_delta} for small
$\bar\chi_0(\rho)$ and small $\delta$, one obtains
\begin{multline}
	\frac{\beta\mathcal F_{v_0}(\rho,\delta)}{N}
	=
	-\rho B_2^{\mathrm{HS}}\,d
	\int_{-\infty}^{\infty}dh\,e^h f_{v_0}(h)
	\\
	\times [
	1
	+
	\frac{\bar\chi_0^{\,2}(\rho)}{2}
	+
	\frac{\bar\chi_0^{\,2}(\rho)}{2}\delta^4
	+
	O (
	\bar\chi_0^{\,3},
	\bar\chi_0^{\,2}\delta^6
	)
	].
	\label{Fv0_chi_delta_expansion}
\end{multline}
Similarly, Eq. \eqref{Fwp_delta} reduces to
\begin{multline}
	\frac{\beta\mathcal F_{w_p}(\rho,\delta)}{N}
	=
	\frac{\beta\mathcal F_{w_p}(\rho,0)}{N}
	-
	\rho B_2^{\mathrm{HS}}d
	\int_{-\infty}^{\infty}dh\,e^h
	\\
	\times \Big[
	a_0(h,\beta\bar u_0,\beta\bar p^{\,2},\nu)
	+
	a_1(h,\beta\bar u_0,\beta\bar p^{\,2},\nu)\,
	\bar\chi_0(\rho)
	\\
	+
	a_2(h,\beta\bar u_0,\beta\bar p^{\,2},\nu)\,
	\bar\chi_0^2(\rho)
	+
	O(\bar\chi_0^3)
	\Big]
	\delta^2
	+
	O(\delta^4),
	\label{Fwp_delta_expansion}
\end{multline}
with
\begin{align}
	a_0
	&=
	e^{-\beta\bar u_0e^{-\nu h}}
	\frac{
		(\beta\bar p^{\,2}e^{-h}\theta(h))^3
	}{
		\big(1-(\beta\bar p^{\,2}e^{-h}\theta(h))^2\big)^{3/2}
	},
	\label{a0_explicit}
	\\[2mm]
	a_1
	&=
	3\nu\beta\bar u_0e^{-\nu h}
	e^{-\beta\bar u_0e^{-\nu h}}
	\frac{
		(\beta\bar p^{\,2}e^{-h}\theta(h))^2
	}{
		\big(1-(\beta\bar p^{\,2}e^{-h}\theta(h))^2\big)^{5/2}
	},
	\label{a1_explicit}
	\\[2mm]
		a_2
	&=
	\frac{3\nu^2}{2}\,
	e^{-\beta\bar u_0e^{-\nu h}}\,
	\beta\bar u_0e^{-\nu h}
	\notag
	\\
	&\quad \times
	\left(
	\beta\bar u_0e^{-\nu h}-1
	\right)
	\frac{
		\beta\bar p^{\,2}e^{-h}\theta(h)
		\left[
		2+3(\beta\bar p^{\,2}e^{-h}\theta(h))^2
		\right]
	}{
		\left[
		1-(\beta\bar p^{\,2}e^{-h}\theta(h))^2
		\right]^{7/2}
	}.
	\label{a2_explicit}
\end{align}
The coefficients entering Eq. \eqref{Fwp_delta_expansion} satisfy $
	a_0>0$, $a_1>0$
whereas no parameter-independent sign can be assigned to
$a_2$. 
Combining Eqs. \eqref{Fid_ubar}, \eqref{Fv0_chi_delta_expansion},
and \eqref{Fwp_delta_expansion}, one obtains, up to quadratic order
in $\delta$ and $\bar\chi_0(\rho)$,
\begin{equation}
	\frac{\beta F(\rho,\delta)}{N}
	=
	\frac{\beta F(\rho,0)}{N}
	+
	d\,A(\rho,\beta)\,\delta^2
	+
	O(\bar\chi_0^3\delta^2)
	+
	O(\delta^4),
	\label{F_Landau}
\end{equation}
where
\begin{multline}
	A(\rho,\beta)
	=
	\frac{1}{2}
	-
	\rho B_2^{\mathrm{HS}}
	\int_{-\infty}^{\infty}dh\,e^h
	\Big[
	a_0(h,\beta\bar u_0,\beta\bar p^{\,2},\nu)
	\\
	+
	a_1(h,\beta\bar u_0,\beta\bar p^{\,2},\nu)\,
	\bar\chi_0(\rho)
	+
	a_2(h,\beta\bar u_0,\beta\bar p^{\,2},\nu)\,
	\bar\chi_0^2(\rho)
	\Big].
	\label{A_rho_beta}
\end{multline}
Notice that $\mathcal F_{v_0}$ does not contribute to the quadratic term in
$\delta$ up to second order in $\bar\chi_0$.
Since $a_0>0$ and $a_1>0$, whereas $a_2$ has no definite sign,
Eq. \eqref{A_rho_beta} shows that $A(\rho,\beta)$ may vanish. Hence,
for fixed $\bar u_0$ and $\bar p^{\,2}$, suitable values of $\beta$ and $\rho$ may satisfy $A(\rho,\beta)=0$, resulting in the onset of a ferroelectric phase transition according to the Landau criterion. Supplementary Note I shows that the possibility of a ferroelectric phase transition, identified by the vanishing of the quadratic coefficient in the Landau free-energy expansion, persists when the leading nonlinear correction in $\Delta_{ij}$ to $\lambda_{ij}$, according to Eq. \eqref{lij_anh}, is included in Eq. \eqref{lambdaij}. The above conclusion assumes that the coefficient of the term $\delta^4$ in the free energy expansion remains positive. The possibility that this coefficient vanishes, possibly leading to the occurrence of a tricritical point and to a first-order ferroelectric phase transition, has been discussed in detail in Ref. \cite{Izzo1} and lies beyond the scope of the present manuscript.

\begin{figure}[t]
	\centering
	\includegraphics[width=6.5cm]{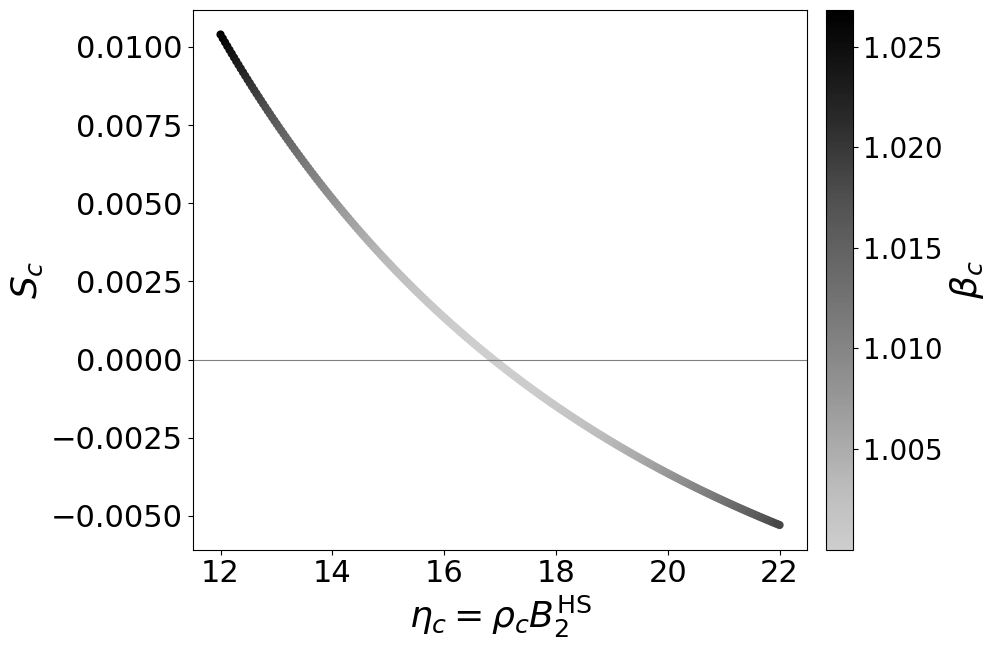}
	\caption{Coefficient $S_c$ along the continuous ferroelectric critical line as a function of the reduced critical density $\eta_c$. The grayscale represents the corresponding value of $\beta_c$.  The critical line is obtained by setting $A(\rho_c,\beta_c)=0$
		in Eq.~\eqref{A_rho_beta}, retaining terms through second order
		in $\bar\chi_0$, with $\bar u_0=24$, $\bar p^{\,2}=0.45$,
		$\nu=2$, $a=0.2$, $b=0.057$, and $\eta_0=16.9$.
		$S_c$ is expressed in units of $dB_2^{\mathrm{HS}}$. Positive (negative) values of $S_c$ correspond to positive (negative) values of $\Delta \rho$.}
	\label{fig:Sc}
\end{figure}

A key question is whether the onset of ferroelectric ordering is accompanied by a transition in the equilibrium density. 
The equation of state in the
isobaric ensemble, Eq. \eqref{EOS_chi_rho}, is evaluated near the ferroelectric critical point
$(\rho,\beta,\delta)=(\rho_c,\beta_c,0)$.
Along the critical isobar, the density of the ferroelectric phase, $\rho=\rho_c+\Delta\rho$, satisfies
\begin{equation}
	P(\rho_c,\beta_c,0)
	=
	P(\rho_c+\Delta\rho,\beta_c,\delta)
	=
	P_c. \label{Pc}
\end{equation}
Expanding Eq. \eqref{Pc} to leading order in $\Delta\rho$ and $\delta^2$
yields
\begin{equation}
	0
	=
	\left.
	\frac{\partial (\beta_c P)}
	{\partial \rho}
	\right|_{\substack{\rho=\rho_c\\ \delta=0}}
	\Delta\rho
	+
	\left.
	\frac{\partial (\beta_c P)}
	{\partial \delta^2}
	\right|_{\substack{\rho=\rho_c\\ \delta=0}}
	\delta^2
	+
	O(\Delta\rho^2,\delta^4).
	\label{EOS_expansion}
\end{equation}
Notice that the scalar quantity $P$ depends on 
$\boldsymbol{\delta}$ only through the rotational invariant
$\delta^2=\boldsymbol{\delta}\!\cdot\!\boldsymbol{\delta}$.
Therefore, the leading dependence of $P$ on $\delta$ is quadratic.
Solving Eq. \eqref{EOS_expansion} for $\Delta\rho$, 
\begin{equation}
	\Delta\rho
	=
	-
	\bigg(
	\left.
	\frac{\partial (\beta_c P)}
	{\partial \rho}
	\right|_{\substack{\rho=\rho_c\\ \delta=0}}
	\bigg)^{-1}
	\left.
	\frac{\partial (\beta_c P)}
	{\partial \delta^2}
	\right|_{\substack{\rho=\rho_c\\ \delta=0}}
	\delta^2
	+
	O(\delta^4).
	\label{density_shift}
\end{equation}
Mechanical stability requires the isothermal compressibility $K_T=
\frac{1}{\rho}
\left(\frac{\partial P}{\partial\rho}\right)_T^{-1}$ to be
positive, or equivalently
$\partial P/\partial\rho>0$. Therefore, Eq. \eqref{density_shift}
shows that the sign of $\Delta \rho$ at the onset of
ferroelectric phase transition is entirely determined by $
\frac{\partial (\beta_c P)}
{\partial (\delta^2)}$ computed at $(\rho, \delta)=(\rho_c,0)$.
Differentiating Eq. \eqref{EOS_chi_rho} with respect to $\delta^2$ at
fixed $\rho$ yields
\begin{equation}
	\left.
	\frac{\partial (\beta_c P)}
	{\partial \delta^2}
	\right|_{\substack{\rho=\rho_c\\ \delta=0}}
	=
	\rho_c^2
	\left.
	\frac{\partial\mathcal B_2}
	{\partial \delta^2}
	\right|_{\substack{\rho=\rho_c\\ \delta=0}}
	+
	\rho_c^3
	\left.
	\frac{\partial^2\mathcal B_2}
	{\partial\bar\chi_0\,\partial \delta^2}
	\right|_{\substack{\rho=\rho_c\\ \delta=0}}
	\left.
	\frac{d\bar\chi_0}{d\rho}
	\right|_{\rho=\rho_c}.
	\label{dP_ddelta2}
\end{equation}
At the critical point of a continuous ferroelectric phase transition,
the $\delta$-quadratic term of the excess free-energy must compensate the increase in the free energy due to
dipole orientational entropy and  must therefore be
negative, implying
\begin{equation}
	\left.
	\frac{\partial\mathcal B_2}
	{\partial \delta^2}
	\right|_{\substack{\rho=\rho_c\\ \delta=0}}
	<0.
\end{equation} 
Hence, the first term in Eq. \eqref{dP_ddelta2}, by itself, drives an
increase of the equilibrium density in the ferroelectric phase.
The second term arises from the density dependence of the
structural compliance $\bar\chi_0$.
For sufficiently small
$\bar\chi_0$, $a_1>0$ implies
\begin{equation}
	\left.
	\frac{\partial^2\mathcal B_2}
	{\partial\bar\chi_0\,\partial\delta^2}
	\right|_c<0.
\end{equation}
Since the constitutive relation assumed for $\bar\chi_0(\rho)$ satisfies
$d\bar\chi_0/d\rho<0$, the second contribution in
Eq. \eqref{dP_ddelta2} opposes the first one and, on its own,
favors a decrease of the equilibrium density upon the onset of
ferroelectric ordering. Eq. \eqref{density_shift} thus shows that the onset of the ferroelectric phase transition is accompanied by a variation of the equilibrium
density. The sign of this variation is not universal, but results from
the competition between the two mechanisms discussed above.
Accordingly, for given values of $\bar u_0$ and $\bar p^{\,2}$, the
equilibrium density may either increase or decrease at the onset of
the ferroelectric phase transition, as a function of $\beta_c$ and $\rho_c$. As shown in Supplementary Note I, the liquid-liquid phase transition accompanying the ferroelectric phase transition persists when the leading nonlinear correction in $\Delta_{ij}$ to $\lambda_{ij}$, as in Eq. \eqref{lij_anh}, is included. Ferroelectric ordering can still be accompanied by either an increase or a decrease in the equilibrium density, depending on the relative magnitude and sign of the density derivatives of the linear and nonlinear contributions to $\lambda_{ij}$, as well as on the thermodynamic conditions and interaction-potential parameters.
To illustrate this behavior, Fig. \ref{fig:Sc} shows,
within the second-order approximation in $\bar\chi_0$,
the coefficient
\begin{equation}
	S_c\equiv
	-\frac{1}{\rho_c^2}
	\left.
	\frac{\partial(\beta_cP)}
	{\partial \delta^2}
	\right|_{\substack{\rho=\rho_c\\ \delta=0}},
	\label{Sc_definition}
\end{equation}
as a function of the reduced critical density
$\eta_c=\rho_c B_2^{\mathrm{HS}}$ along the ferroelectric critical line, i.e. the locus of points satisfying
$A(\rho_c,\beta_c)=0$, see Eq. \eqref{F_Landau}. Unless otherwise stated, all calculations were carried out with
\begin{equation}
	\bar u_0=1,\qquad
	\bar p^{\,2}=0.45,\qquad
	\nu=2, \label{parameters}
\end{equation}
together with the following exponential constitutive relation for
$\bar\chi_0(\eta)$, with $\eta=\rho B_2^{HS}$
\begin{equation}
	\bar\chi_0(\eta)=a\,e^{-b(\eta-\eta_0)}. \label{chi0exp}
\end{equation}
According to Eq. \eqref{density_shift}, $S_c$ determines the sign of $\Delta \rho$ and, together with the isothermal compressibility,  quantifies its magnitude for a given $\delta^2$. Fig. \ref{fig:Sc} shows that, for the parameters considered therein,
ferroelectric order favours an increase in equilibrium
density at lower $\eta_c$ and a decrease at higher
$\eta_c$.
If $d\bar\chi_0/d\rho>0$, the two contributions in Eq. \eqref{dP_ddelta2} have the same sign and both favor an increase in the equilibrium density upon the onset of ferroelectric ordering, provided that $\partial^2\mathcal B_2/(\partial\bar\chi_0\,\partial\delta^2)<0$, i.e. for sufficiently small $\bar\chi_0$. Thus, in this regime, ferroelectric ordering would always be accompanied by an increase in the equilibrium density.

\begin{figure*}[t]
	\centering
	\includegraphics[width=14.0cm]{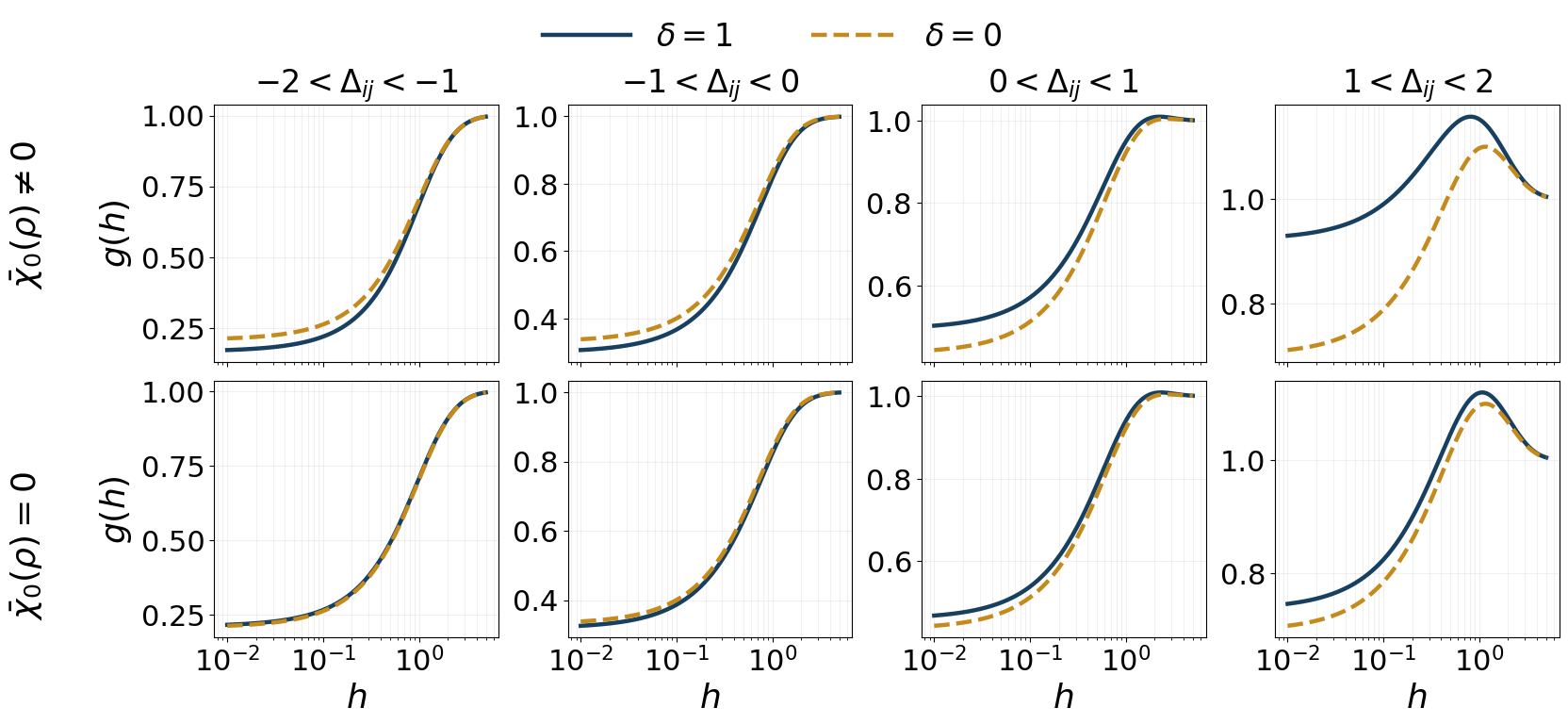}
	\caption{
		Conditional pair correlation functions $g(h\,|\,\Delta_{ij} \in I)$ for four intervals of $\Delta_{ij}$.
		Solid (dashed) lines denote ferroelectric, $\delta=1$ (paraelectric, $\delta=0$) state.
		The upper panels show $g(h\,|\,\Delta_{ij})$ obtained with $\bar{\chi}_0(\rho)\neq0$, whereas the lower panels show the same quantity obtained by setting $\bar{\chi}_0=0$ keeping $(\eta,\beta,\delta)$ fixed for each state. The coarse-grained theory parameters and reference states are those specified in the text and are also used in Figs. \ref{Fig2_bis_Pdelta} and \ref{Fig2_bis_grtot}.
	}
	\label{fig:conditional_gr}
\end{figure*}

It is analysed in the following how the ferroelectric phase transition and the associated
liquid density variation encodes signatures in the radial distribution function
$g(r)$.
To isolate the structural changes induced
by ferroelectric ordering, $g(r)$ is evaluated for two states at the
same $T$ and $P$: a paraelectric state ($\delta=0$) and a ferroelectric reference state ($\delta=1$). 
Since the aim is to characterize the structural changes associated with
the onset of ferroelectric phase transition, the common $T$ and $P$ 
are chosen equal to $T_c=1/(k_B\beta_{c})$ and $P_c$. 
The parameters of Eq. \eqref{chi0exp} are set to
$a=0.15$, $b=2.2$, and $\eta_0=14$.
The reference states used in Figs. \ref{fig:conditional_gr}-\ref{fig:conditional_gr} are characterized by the paraelectric and ferroelectric reduced densities $\eta_P=16$, $\eta_F=14.26$, and
$\beta_c=0.97$, corresponding to the paraelectric and ferroelectric structural compliances
$\bar\chi_0^P=1.84 \ 10^{-3}$ and
$\bar\chi_0^F=8.48 \ 10^{-2}$.
The relative density variation is
$\Delta\rho/\rho_P\simeq-0.11$.
The numerical procedure used to obtain the densities of the two states is described in Appendix \eqref{AppendixI}.

\begin{figure*}[t]
	\centering
	\includegraphics[width=14.5cm]{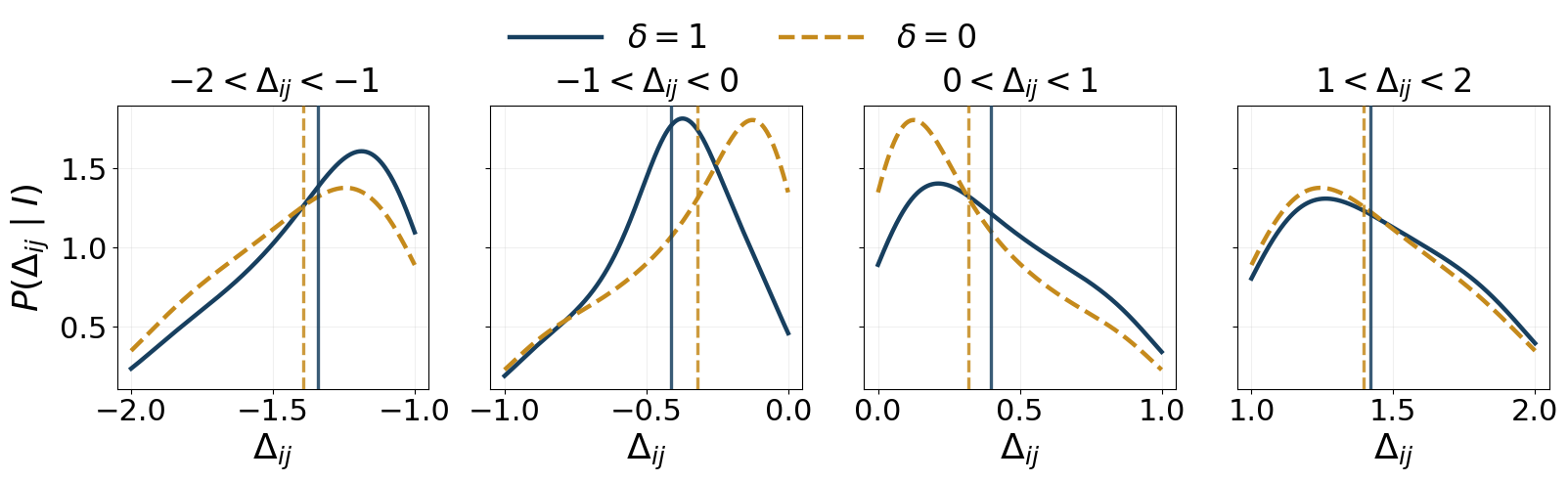}
	\caption{
		Probability distributions
		$P(\Delta_{ij}| I)$ for the four intervals $I$, obtained numerically by Gaussian kernel density estimation. Solid (dashed) lines denote ferroelectric, $\delta=1$ (paraelectric, $\delta=0$) state. Vertical lines mark the corresponding centroids. The coarse-grained theory parameters and reference states are the same as in Fig. \ref{fig:conditional_gr}.
	}
	\label{Fig2_bis_Pdelta}
\end{figure*}
Given the identity $\Delta_{ij}
=
\tilde\theta_i\tilde\theta_j
-
\hat d_i\!\cdot\!\hat d_j$, the pair correlation functions conditioned on
$\Delta_{ij}\in I$,
\begin{multline}
	g(h|\Delta_{ij} \in I)
	= \frac{1}{Z(\Delta_{ij} \in I)}
	\displaystyle
	\int_{\Delta_{ij}(\tilde{\mathbf t}_{ij}) \in I
	}
	d\tilde{\mathbf t}_{ij}\,
	\\
	\times
	\frac{
		e^{-\frac{1}{2}
			\tilde{\mathbf t}_{ij}
			\mathbf G^{-1}
			\tilde{\mathbf t}_{ij}^{T}}
	}{
		2\pi(\det\mathbf G)^{1/2}
	}
	e^{-\beta\left[
		v_0(h,\Delta_{ij})
		+w_p(h,\Delta_{ij})
		\right]},
	\label{eq:g_conditional_num}
\end{multline}
with 
\begin{equation}
		\displaystyle
		Z(\Delta_{ij}\in I)=\int_{\Delta_{ij}(\tilde{\mathbf t}_{ij}) \in I}
		d\tilde{\mathbf t}_{ij}\,
		\frac{
			e^{-\frac{1}{2}
				\tilde{\mathbf t}_{ij}
				\mathbf G^{-1}
				\tilde{\mathbf t}_{ij}^{T}}
		}{
			2\pi(\det\mathbf G)^{1/2}
		} \label{ZdeltaijinI}
\end{equation}
were evaluated numerically via Gauss-Hermite quadrature. To derive Eq. \eqref{eq:g_conditional_num}, Eq. \eqref{g_dinf} has
been used. 
The four intervals
\begin{multline}
	I_1=(-2,-1), \qquad
	I_2=(-1,0),\\
	I_3=(0,1), \qquad
	I_4=(1,2).
	\label{intervals}
\end{multline}
were considered. For comparison, the corresponding $g(h|\Delta_{ij}\in I)$ for
$\bar{\chi}_0=0$, which isolate the structural signatures of the bare ferroelectric phase transition,  were
also computed. 
Fig. \ref{fig:conditional_gr} shows $g(h|\Delta_{ij}\in I)$ for
$\bar{\chi}_0(\rho)\neq0$ and $\bar{\chi}_0(\rho)=0$.
Consider first the latter case. For $\Delta_{ij}>0$, the ferroelectric
phase exhibits a smaller effective pair contact distance than the
paraelectric phase, as indicated by the shift of the inflection point
of $g(h|\Delta_{ij}\in I)$. For $\Delta_{ij}<0$, instead,
the shift is much weaker: the effective pair contact distance is
slightly reduced for $-1<\Delta_{ij}<0$ and remains essentially
unchanged for $-2<\Delta_{ij}<-1$.
Since $v_0(h)$ is the same in the paraelectric and ferroelectric
states for $\bar{\chi}_0=0$, the differences in
$g(h|\Delta_{ij}\in I)$ therefore arise from averaging the same dipolar Boltzmann factor $e^{-\beta w_p(h, \Delta_{ij})}$ over different distribution of $\Delta_{ij}$ within each interval
\begin{multline}
	P(\Delta_{ij}|I)
	=
	\frac{1}{Z(\Delta_{ij}\in I)}
	\int_{\Delta_{ij}(\tilde{\mathbf t}_{ij}) \in I} d\tilde{\mathbf t}_{ij}\,
	\\
	\times
	\frac{
		e^{-\frac{1}{2}
			\tilde{\mathbf t}_{ij}
			\mathbf G^{-1}
			\tilde{\mathbf t}_{ij}^{T}}
	}{
		2\pi(\det\mathbf G)^{1/2}
	}
	\delta\!\left(
	\Delta_{ij}
	-\tilde\theta_i\tilde\theta_j
	+\hat d_i\cdot\hat d_j
	\right),
	\label{PDeltaI}
\end{multline}
which is shown in Fig. \ref{Fig2_bis_Pdelta} for the four intervals defined in
Eq. \eqref{intervals}. For graphical representation, the probability
densities are reconstructed
numerically using
Gaussian kernel density estimation. The
centroids are computed before smoothing.
The figure shows that
ferroelectric ordering modifies $P(\Delta_{ij}|I)$, shifting its centroid
slightly towards less negative values for $-2<\Delta_{ij}<-1$,
	and
towards more negative values for
$-1<\Delta_{ij}<0$
and towards more positive values for the intervals $\Delta_{ij}>0$.
For fixed $h$ and $\Delta_{ij}$, the dipolar Boltzmann factor
$e^{\beta\bar p^{\,2}\Delta_{ij}e^{-h}}$ entering
$g(h|\Delta_{ij}\in I)$ is the same in the paraelectric and
ferroelectric states. At fixed $h$, this factor increases monotonically
with $\Delta_{ij}$, whereas, for fixed $\Delta_{ij}>0$, it decreases
with increasing $h$ and approaches unity as $h\rightarrow\infty$.
Consequently, a shift of $P(\Delta_{ij}\mid I)$ within a given interval towards larger positive
values of $\Delta_{ij}$ enhances the contribution of short $h$
to $g(h|\Delta_{ij}\in I)$, whereas a shift
towards more negative values suppresses it. The two effects are
asymmetric, since
$
\frac{\partial}{\partial\Delta_{ij}}e^{-\beta w_p}
=
\beta\bar p^{\,2}e^{-h}e^{-\beta w_p}
$
increases exponentially with $\Delta_{ij}$. Thus, for the equal shift in $\Delta_{ij}$, the variation of the
dipolar Boltzmann factor is larger towards positive than towards
negative $\Delta_{ij}$.

If $\bar{\chi}_0(\rho)\neq0$, the density decrease accompanying
ferroelectric ordering introduces an additional mechanism affecting
$g(h|\Delta_{ij}\in I)$. Since $\eta_F<\eta_P$, the constitutive
relation implies $\bar{\chi}_0^F>\bar{\chi}_0^P$. According to
$\beta v_0(h,\Delta_{ij})
=
\beta\bar u_0
e^{-\nu h-\nu\bar{\chi}_0(\rho)\Delta_{ij}}$, an increase of $\bar{\chi}_0$ decreases $v_0$ for
$\Delta_{ij}>0$, whereas it increases $v_0$ for
$\Delta_{ij}<0$. Since $v_0$ decays exponentially with $h$, these effects are
confined to small $h$. 
Consequently, through the Boltzmann factor
$e^{-\beta v_0(h,\Delta_{ij})}$ entering
$g(h|\Delta_{ij}\in I)$, pair configurations at small $h$ are assigned a larger statistical weight for
$\Delta_{ij}>0$ and a smaller one for $\Delta_{ij}<0$. As a result, for $\Delta_{ij}>0$, introducing $\bar \chi_0(\rho) \neq 0$ reinforces the short-distance enhancement of
$g(h|\Delta_{ij}\in I)$ already observed for $\bar \chi_0=0$.
For $\Delta_{ij}<0$, instead, it suppresses short-distance contribution to $g(h|\Delta_{ij}\in I)$
more strongly in the ferroelectric phase, producing a clear separation
from the paraelectric $g(h|\Delta_{ij}\in I)$ that is absent for
$\bar{\chi}_0=0$.
Fig. \ref{fig:conditional_gr} thus shows how the effective contact distance
$\lambda_{ij}$ is encoded in $g(h|\Delta_{ij}\in I)$. For
$\Delta_{ij}>0$, the smaller $\lambda_{ij}$ shifts the onset of the
rise in $g(h|\Delta_{ij}\in I)$ towards shorter interparticle
distances.
Conversely, for $\Delta_{ij}<0$, the larger $\lambda_{ij}$ shifts this
onset towards larger distances.

\begin{figure}[t]
	\includegraphics[width=4.7cm]{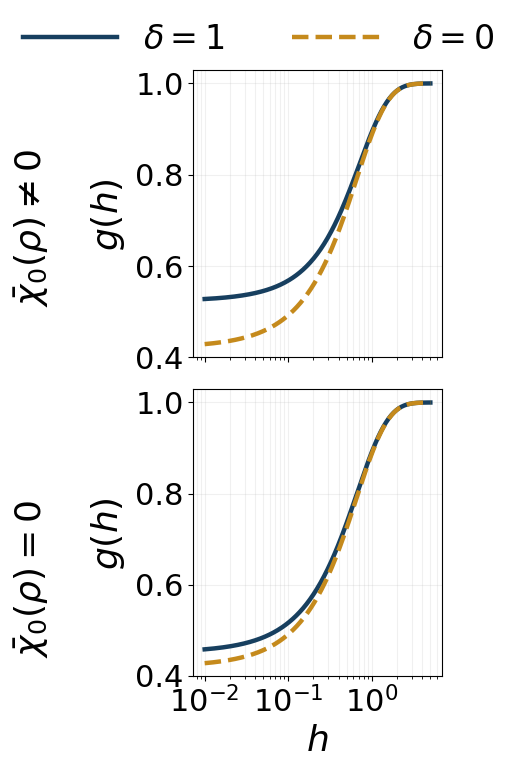}
	\caption{
		Total pair correlation function $g(h)$.
		Solid (dashed) lines correspond to the ferroelectric, $\delta=1$ (paraelectric, $\delta=0$) phase.
		Upper (lower) panels correspond to $\bar{\chi}_0(\rho)\neq0$ ($\bar{\chi}_0=0$). For each state, the same values of $(\eta,\beta,\delta)$ are used. The coarse-grained theory parameters and reference states are the same as in Fig. \ref{fig:conditional_gr}.}
	\label{Fig2_bis_grtot}
\end{figure}

In the case of the total pair correlation functions shown in
Fig. \ref{Fig2_bis_grtot}, the enhancement at small $h$ for
$\Delta_{ij}>0$ is partially compensated by the suppression for
$\Delta_{ij}<0$. As a result, the change in the orientationally
averaged $g(h)$ remains relatively small, with the residual enhancement
in the ferroelectric phase reflecting the incomplete compensation
between the two contributions.
This is a distinctive consequence of the
$\Delta_{ij}$-dependent effective pair contact distance
$\lambda_{ij}$. Although the ferroelectric phase is characterized by a
lower density and a larger effective excluded volume these changes
are not directly reflected in the total pair correlation function. By contrast, an
isotropic increase of the excluded volume, affecting all particle pairs
equally, would produce a clear shift of the first-neighbour structure
in the total $g(h)$.

\subsection{A Liquid-Liquid Phase Transition within the Paraelectric Phase} \label{SecIIb}
One asks whether a first-order
liquid-liquid phase transition can also occur entirely within the
paraelectric phase ($\delta=0$), where the dipole orientational
distribution function remains isotropic,
\begin{equation}
	\zeta(\hat d)=\frac{1}{\Omega_d},
	\label{zetaiso}
\end{equation}
and whether the corresponding coexistence line can terminate at a
second-order critical point.
Eq. \eqref{EOS_chi_rho} is indeed nonlinear in $\rho_{eq}$ and may
therefore admit multiple solutions at fixed pressure, a necessary
condition for a first-order phase transition.
To determine whether a first-order phase transition does occur
within the paraelectric phase, the limit of weak
$\bar\chi_0(\rho)$ is considered and $\mathcal B_2$ in
Eq. \eqref{EOS_chi_rho} is expanded about $\bar\chi_0=0$ at $\delta=0$. The transition is then investigated numerically for a specific choice of parameters.
Using Eqs. \eqref{F_v0_h}, \eqref{F_wp_h}, and
\eqref{phibar_chi_series1}, one obtains
\begin{multline}
	\mathcal B_2(\beta,\bar\chi_0)
	=
	-dB_2^{\mathrm{HS}}
	\int_{-\infty}^{\infty}dh\,e^h
 [
	b_0(h,\beta\bar u_0,\beta\bar p^{\,2},\nu)
	\\
	+
	b_1(h,\beta\bar u_0,\beta\bar p^{\,2},\nu)\,
	\bar\chi_0
	+
	b_2(h,\beta\bar u_0,\beta\bar p^{\,2},\nu)\,
	\bar\chi_0^2
	+
	O(\bar\chi_0^3)
	],
	\label{eq:B2_para_expansion}
\end{multline}
with
\begin{align}
	b_0
&=
e^{-\beta\bar u_0e^{-\nu h}}-1
\notag\\
&\quad+
e^{-\beta\bar u_0e^{-\nu h}}
\bigg[
\frac{1}{
	\sqrt{
		1-(\beta\bar p^{\,2}e^{-h}\theta(h))^2
	}
}
-1
\bigg],
\label{b0_explicit}
	\\[2mm]
	b_1
	&=
	\nu\beta\bar u_0e^{-\nu h}
	e^{-\beta\bar u_0e^{-\nu h}}
	\frac{
		\beta\bar p^{\,2}e^{-h}\theta(h)
	}{
		\left[
		1-(\beta\bar p^{\,2}e^{-h}\theta(h))^2
		\right]^{3/2}
	},
	\label{b1_explicit}
	\\[2mm]
	b_2
&=
\frac{1}{2}
(
e^{-\beta\bar u_0e^{-\nu h}}-1
)
+
\frac{\nu^2}{2}\,
\beta\bar u_0e^{-\nu h}
\left(
\beta\bar u_0e^{-\nu h}-1
\right)
\notag\\
&\quad\times
e^{-\beta\bar u_0e^{-\nu h}}
\bigg[
\frac{
	1+2(\beta\bar p^{\,2}e^{-h}\theta(h))^2
}{
	[
	1-(\beta\bar p^{\,2}e^{-h}\theta(h))^2
	]^{5/2}
}
-1
\bigg].
\label{b2_explicit}
\end{align}
The coefficient $ b_1>0$.
Neither $b_0$ nor $b_2$ has a definite sign.
Differentiating Eq. \eqref{eq:B2_para_expansion} with respect to
$\bar\chi_0$ gives
\begin{multline}
	\frac{\partial\mathcal B_2}
	{\partial\bar\chi_0}
	=
	-dB_2^{\mathrm{HS}}
	\int_{-\infty}^{\infty}dh\,e^h
\left[
	b_1(h,\beta\bar u_0,\beta\bar p^{\,2},\nu)
	\right.
	\\
	\left.
	+
	2b_2(h,\beta\bar u_0,\beta\bar p^{\,2},\nu)\bar\chi_0
	+
	O(\bar\chi_0^2)
	\right].
	\label{eq:dB2_dchi_para}
\end{multline}
Substitution of Eqs. \eqref{eq:B2_para_expansion} and
\eqref{eq:dB2_dchi_para} into Eq. \eqref{EOS_chi_rho} yields, 
\begin{multline}
	B_2^{\mathrm{HS}}\beta P
	=
	\eta
	-
	d\eta^2
	\int_{-\infty}^{\infty}dh\,e^h\Big[
	b_0(h,\beta\bar u_0,\beta\bar p^{\,2},\nu)
	\\
	+
	b_1(h,\beta\bar u_0,\beta\bar p^{\,2},\nu)\bar\chi_0
	+
	b_2(h,\beta\bar u_0,\beta\bar p^{\,2},\nu)\bar\chi_0^2
	\\
	+
	\eta
	\left[
	b_1(h,\beta\bar u_0,\beta\bar p^{\,2},\nu)
	+
	2b_2(h,\beta\bar u_0,\beta\bar p^{\,2},\nu)\bar\chi_0
	\right]
	\frac{d\bar\chi_0}{d\eta}
	\Big]\\
	+
	O(\bar\chi_0^3).
	\label{eq:EOS_perturbative_general}
\end{multline}
It is convenient to introduce the reduced
pressure
\begin{equation}
	\Pi
	=
	\frac{\beta P B_2^{\mathrm{HS}}}{d},
	\label{eq:Pi_def}
\end{equation}
which remains finite in the large-$d$ limit.  
The reduced Helmholtz free energy per particle and the reduced chemical
potential are consistently defined as
\begin{equation}
	\varphi(\eta,\beta)
	=
	\frac{1}{d}\frac{\beta F}{N},
	\qquad
	\mathcal M(\eta,\beta)
	=
	\frac{\beta\mu}{d},
\end{equation}
where $\mu$ denotes the chemical potential.
Substituting Eq. \eqref{eq:B2_para_expansion} into
Eq. \eqref{F_isobaric_large_d}, the reduced free energy reads
\begin{multline}
	\varphi(\eta,\beta)
	=
	\frac{1}{d}
	\Big[
	\log\!\Big(
	\frac{\eta\Lambda^d}{B_2^{\mathrm{HS}}}
	\Big)
	-1
	-\log\Omega_d
	\Big]
	\\
	-
	\eta
	\int_{-\infty}^{\infty}dh\,e^h
	\left[
	b_0
	+
	b_1\bar\chi_0(\eta)
	+
	b_2\bar\chi_0^2(\eta)
	\right]
	+
	O(\bar\chi_0^3).
	\label{eq:f_pert_LL}
\end{multline}
In the limit $d\to\infty$ the reduced
density-dependent contribution to $F_{id}$
vanishes. 
The chemical potential and pressure, together with their reduced counterparts, satisfy the thermodynamic identities
\begin{equation}
	\mu
	=
	\frac{F}{N}
	+
	\rho
	\frac{\partial(F/N)}{\partial\rho},
	\qquad
	\mathcal M
	=
	\varphi
	+
	\eta
	\frac{\partial\varphi}{\partial\eta},
	\label{eq:M_def}
\end{equation}
and
\begin{equation}
	P
	=
	\rho^2
	\frac{\partial(F/N)}{\partial\rho},
	\qquad
	\Pi
	=
	\eta^2
	\frac{\partial\varphi}{\partial\eta}.
	\label{eq:Pi_phi}
\end{equation}
In the numerical calculations, both the reduced pressure
$\Pi(\eta,\beta)$ and the reduced chemical potential
$\mathcal M(\eta,\beta)$ are evaluated from the reduced Helmholtz
free energy through Eqs. \eqref{eq:M_def} and
\eqref{eq:Pi_phi}.
The parameters entering the structural compliance $\bar\chi_0(\eta)$ in
Eq.~\eqref{chi0exp} are 
\begin{equation}
	a=0.15,
	\qquad
	b=2.2,
	\qquad
	\eta_0=14,
\end{equation}
so that the perturbative expansion of
Eq. \eqref{eq:B2_para_expansion} remains valid throughout the
density range of interest while retaining a significant density
dependence. The inverse temperature is fixed at the representative value
$\beta^\star=0.880$. At $\beta^\star$, the spinodal reduced densities
are obtained from
\begin{equation}
	\left.
	\frac{\partial\Pi}{\partial\eta}
	\right|_{\beta^\star}
	=0.
\end{equation}
yielding
\begin{equation}
	\eta_{s1}=14.593,
	\qquad
	\eta_{s2}=14.946.
	\label{eta_s}
\end{equation}
For $\eta_{s,1}\le \eta \le \eta_{s,2}$,
$A(\eta,\beta^\star)>0$, so that the dipolar liquid remains
paraelectric throughout the thermodynamic region considered, consistently
with Eq. \eqref{zetaiso}.
Since $\Pi(\eta,\beta^*)$ is a continuous function of $\eta$ and
admits only the two spinodal densities in
Eq. \eqref{eta_s}, for any reduced pressure
$\Pi_0$ satisfying $\Pi_{s,2}<\Pi_0<\Pi_{s,1}$,
where
$\Pi_{s,1(2)}=\Pi(\eta_{s,1(2)},\beta^*)$, the equation
\begin{equation}
	\Pi(\eta,\beta^*)=\Pi_0,
\end{equation}
admits three distinct solutions,
\begin{equation}
	\eta_1<\eta_2<\eta_3.
\end{equation}
The low- and high-density solutions, $\eta_L \equiv \eta_1$ and $\eta_H \equiv\eta_3$, satisfy
$\left.\dfrac{\partial\Pi}{\partial\eta}\right|_{\beta^\star}>0$, 
and are therefore mechanically stable, whereas $\eta_2$ satisfies $\left.\dfrac{\partial\Pi}{\partial\eta}\right|_{\beta^\star}<0$
and is mechanically unstable.
Away from coexistence, only one of the two states, $\eta_L$ and $\eta_H$, is thermodynamically stable, while the other is metastable.
A first-order liquid-liquid phase transition occurs at the pressure
$\Pi^*\in(\Pi_{s,2},\Pi_{s,1})$,
where the two phases satisfy 
\begin{equation}
	\mathcal M(\eta_L,\beta^*)
=
\mathcal M(\eta_H,\beta^*),	\qquad
\Pi(\eta_L,\beta)
=
\Pi(\eta_H,\beta)
=
\Pi^*,
\label{eq:LL_coexistence}
\end{equation}
or, equivalently, have the same Gibbs free energy per particle.
The reduced coexistence densities $\eta_L$ and $\eta_H$, and coexistence pressure $\Pi^\star$, satisfy the Maxwell
equal-area construction,
\begin{equation}
	\int_{\eta_L}^{\eta_H}
	\frac{\Pi(\eta,\beta^\star)-\Pi^\star}
	{\eta^2}\,
	d\eta
	=
	0.
\end{equation}
The coexistence pressure $\Pi^\star$ is determined numerically by locating
the zero of $
\Delta\mathcal{M}(\Pi)
=
\mathcal{M}(\eta_H,\beta^\star)
-
\mathcal{M}(\eta_L,\beta^\star)$,
within the pressure interval
$\Pi_{s,2}<\Pi<\Pi_{s,1}$.
The corresponding coexistence densities $\eta_L$ and $\eta_H$ are then
obtained as the low- and high-density solutions of
$\Pi(\eta,\beta^\star)=\Pi^\star$.
\begin{figure*}[t]
	\centering
	\includegraphics[width=15.0cm]{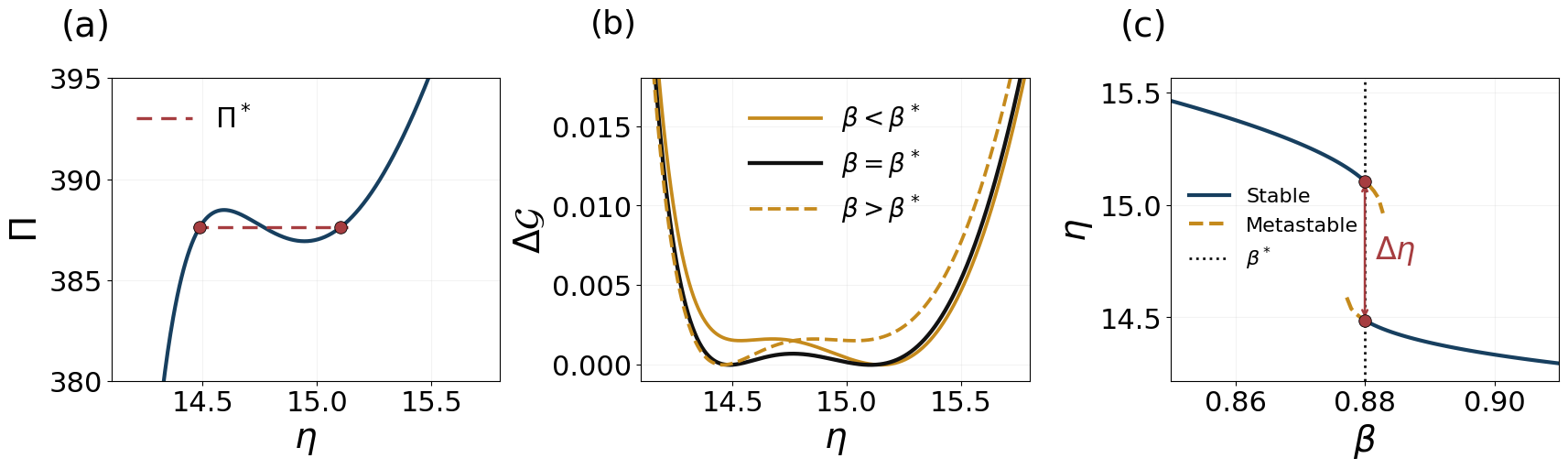}
\caption{
	Numerical evidence for a first-order liquid-liquid phase transition within the
	paraelectric phase at the representative inverse temperature
	$\beta^\star$. (a) Equation of state
	$\Pi(\eta,\beta^\star)$, showing the coexistence pressure
	$\Pi^\star$ and the corresponding coexistence densities
	$\eta_L$ and $\eta_H$ satisfying the Maxwell equal-area
	construction. (b) Reduced Gibbs free energy difference,
	$\Delta\mathcal G=\mathcal G-\min_{\eta}\mathcal G$,
	for $\beta_-=0.878$, $\beta^\star=0.880$, and
	$\beta_+=0.882$. (c) Equilibrium reduced density as a function
	of $\beta$. Solid curves show stable states, dashed curves show metastable states, and the vertical dotted line marks $\beta^*$.
} \label{fig:LLPT_paraelectric}
\end{figure*}
Fig. \ref{fig:LLPT_paraelectric} shows the thermodynamic signatures of
the first-order liquid-liquid phase transition at the representative inverse
temperature $\beta^\star$, within the paraelectric phase. Panel (a)
displays the curve representing the reduced equation of state derived from
Eqs. \eqref{eq:f_pert_LL} and \eqref{eq:Pi_phi}. The Maxwell equal-area
construction identifies the coexistence pressure $\Pi^\star$, and densities $\eta_L$ and $\eta_H$. The
corresponding reduced Gibbs free energy,
\begin{equation}
	\mathcal G(\eta;\beta,\Pi)
	\equiv \varphi(\eta,\beta)+
	\frac{\Pi}{\eta},
	\label{eq:G_reduced}
\end{equation}
is shown in panel (b). At $\beta=\beta^\star$, the two minima are
degenerate, whereas for $\beta\neq\beta^\star$ only one minimum
corresponds to the thermodynamically stable phase, while the other
identifies a metastable state. Panel (c) displays the discontinuous
density jump
\begin{equation}
	\Delta\eta=\eta_H-\eta_L,
\end{equation}
at $\beta^\star$, providing the characteristic signature of a
first-order liquid-liquid phase transition.
\begin{figure*}[t]
	\includegraphics[width=10.0cm]{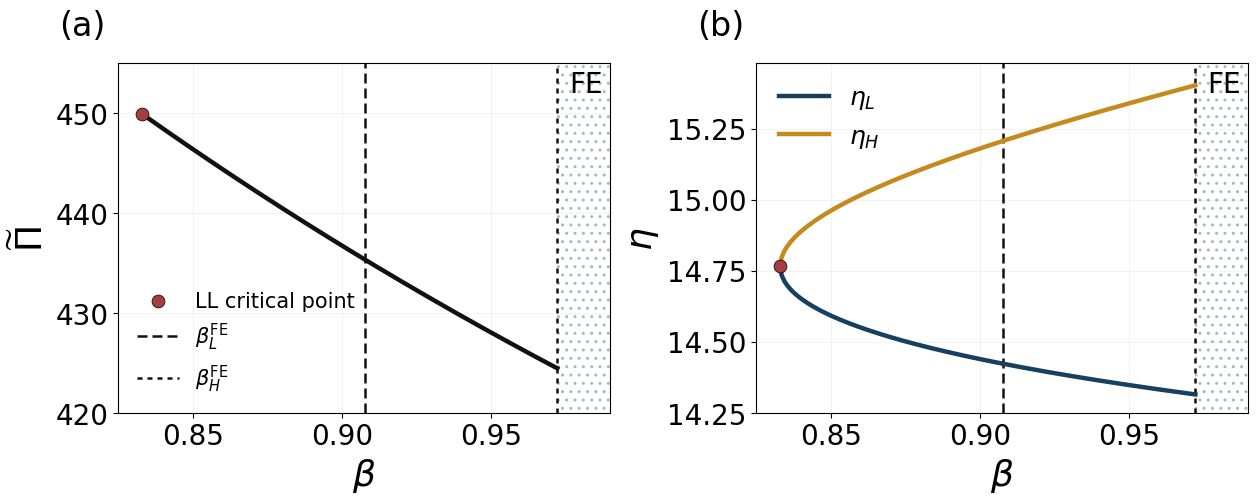}
\caption{
	Liquid-liquid phase diagram. (a) Coexistence line in the
	$(\beta,\widetilde{\Pi})$ plane, with
	$\widetilde{\Pi}\equiv \Pi/\beta=P B_2^{\mathrm{HS}}/d$,
	terminating at the liquid-liquid critical point
	$(\beta_c^{(\rho)},\widetilde{\Pi}_c^{(\rho)})$.
	The vertical dashed and dotted lines mark the onset of ferroelectric instability along the low- and high-density branches, respectively.
}
	\label{fig:critical_density}
\end{figure*}

The analysis is then extended beyond the representative state point
$\beta^\star$ to determine the liquid-liquid coexistence line and its
critical point.
For each $\beta$, the coexistence pressure
$\Pi^*(\beta)$ and the corresponding densities
$\eta_L(\beta)$ and $\eta_H(\beta)$ are determined numerically as
described above. The liquid-liquid critical point is identified by
the conditions
\begin{equation}
	\left.
	\frac{\partial\Pi}{\partial\eta}
	\right|_{\beta}
	=
	0,
	\qquad
	\left.
	\frac{\partial^2\Pi}{\partial\eta^2}
	\right|_{\beta}
	=
	0.
\end{equation}
yielding
\begin{equation}
	\beta_c^{(\rho)}=0.833,
	\qquad
	\eta_c^{(\rho)}=14.765,
	\qquad
	\Pi_c^{(\rho)}=374.8.
\end{equation}
The phase diagram is shown in
Fig. \ref{fig:critical_density}. Panel (a) displays the liquid-liquid
coexistence line in the $(\beta,\widetilde\Pi)$ plane and the related critical point, whereas panel (b) shows
$\eta_L$ and $\eta_H$ as functions of $\beta$. As the critical point is approached, the density jump
$\Delta\eta$ continuously vanishes at
$\beta_c^{(\rho)}$, consistent with the critical endpoint of a
first-order liquid-liquid phase transition.
The dashed lines
mark the onset of the ferroelectric instability of the low- and
high-density branches, determined by the condition $A(\eta_{L(H)},\beta)=0$.
For $\beta>\beta_H^{\rm FE}$, the high-density liquid, which is the
thermodynamically stable phase on the high-temperature side of the
liquid-liquid phase transition, becomes  unstable towards ferroelectric
order ($A<0$). Consequently, the isotropic $\zeta(\hat d)$,
Eq. \eqref{zetaiso}, and the following framework are no longer valid. In the temperature interval
$\beta_L^{\rm FE}<\beta<\beta_H^{\rm FE}$,
an interesting phenomenon emerges, as discussed below.

\begin{figure}[t]
	\centering
	\includegraphics[width=5.7cm]{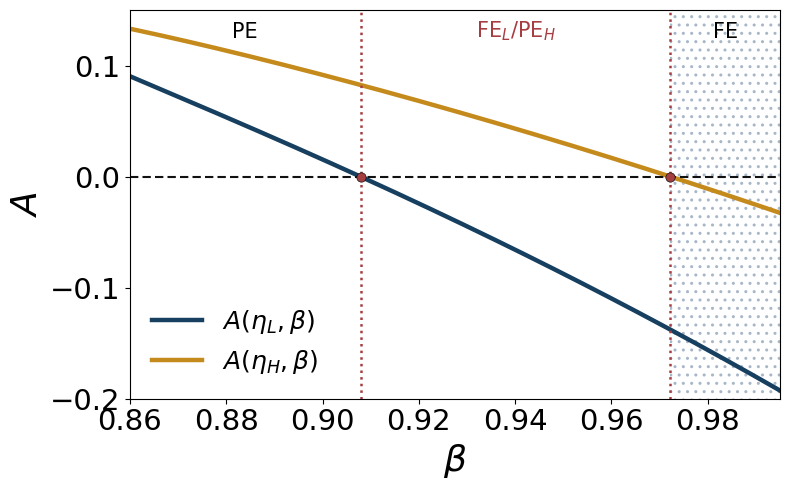}
\caption{
	Landau coefficient $A(\eta,\beta)$ evaluated along the liquid-liquid
	coexistence line for the low-density ($\eta_L$) and high-density
	($\eta_H$) phases. The horizontal dashed line marks the ferroelectric instability threshold, $A=0$, while the vertical lines denote the inverse
	temperatures $\beta_L^{\rm FE}$ and $\beta_H^{\rm FE}$ at which the
	paraelectric solution loses stability on the low- and high-density
	phases, respectively. In the interval
	$\beta_L^{\rm FE}<\beta<\beta_H^{\rm FE}$, the high-density liquid
	remains paraelectric whereas the low-density liquid is unstable towards ferroelectric order.
}
	\label{fig:LL_FE}
\end{figure}

As shown in Fig. \ref{fig:LL_FE}, there exists an interval of $\beta$ for which the high-density liquid is paraelectric
($A>0$), so that the present framework remains applicable. Upon crossing
the first-order liquid-liquid transition line, the Landau coefficient
becomes negative ($A<0$), indicating that the stable low-density phase is
unstable towards ferroelectric order. Hence, in this regime the liquid-liquid phase transition may induce a first-order ferroelectric phase transition.
It should be stressed, however, that this possibility is identified
only a posteriori and is not explicitly incorporated into the derivations of the present section. Indeed, the liquid-liquid coexistence line shown in Fig. \ref{fig:critical_density} 
has been determined throughout by imposing $\zeta(\hat d)$ isotropic as in Eq. \eqref{zetaiso}. 

From Eq. \eqref{EOS_chi_rho},
a nonmonotonic equation of state requires
\begin{equation}
	1+2\rho\mathcal B_2
	+4\rho^2\frac{d\mathcal B_2}{d\rho}
	+\rho^3\frac{d^2\mathcal B_2}{d\rho^2}<0
\end{equation}
over a finite density interval. This condition is not precluded
a priori either by $\bar\chi_0<0$ or by
$d\bar\chi_0/d\eta>0$, since it depends on the full density dependence
of $\mathcal B_2$. Changing either sign can therefore shift, or even eliminate, the spinodal and
coexistence regions and the associated critical point, but does not
exclude their occurrence on general grounds.
Similarly, as shown in Supplementary Note I, the possible occurrence of a liquid-liquid phase transition within the paraelectric phase is not contingent on the absence of nonlinear terms in $\Delta_{ij}$ in $\lambda_{ij}$.

\subsection{On a Microscopic Origin of Two-Scale Jagla Potentials in Dipolar Liquids} \label{SecII3}
\begin{figure}[t]
	\centering
	\includegraphics[width=6.0cm]{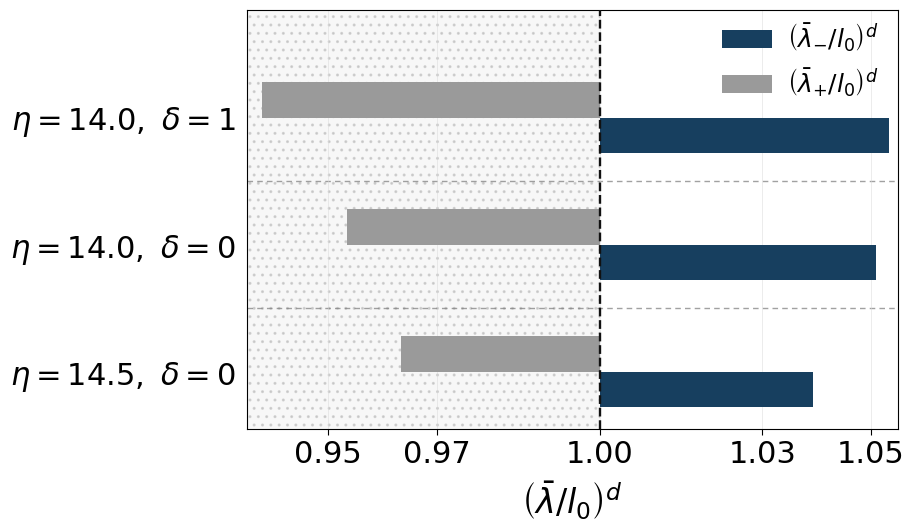}
	\caption{Values of $(\bar \lambda_-/l_0)^d$ and $(\bar \lambda_+/l_0)^d$ for  three representative
		thermodynamic states. The characteristic length $(\bar \lambda_-$ and $(\bar \lambda_+$ are obtained by averaging over
		$\Delta_{ij}\in I_{-}=(-2,0)$ and
		$\Delta_{ij}\in I_{+}=(0,2)$, respectively, according to Eq. \eqref{lambdabar}. The dashed line marks the reference value
		$(\bar{\lambda}/l_0)^d=1$, while the shaded region corresponds to
		$\bar{\lambda}<l_0$.
	}	\label{fig:TwoScaleLength}
\end{figure}

The dependence of 
$\lambda_{ij}$ on 
$\Delta_{ij}$ given in Eq. \eqref{Deltaijd}, following Eq. \eqref{lambdaij}, has an interesting physical consequence. It generates a
family of soft-core potentials of the form given in
Eq. \eqref{v0softlambdaij}, characterized by different effective length scales.
With $\bar \chi_0>0$, $\Delta_{ij}>0$, corresponding to attractive dipolar interaction, gives $\lambda_{ij}<l_0$. Conversely, for
$\Delta_{ij}<0$, corresponding to repulsive dipolar interaction, $\lambda_{ij}>l_0$.
The two classes of soft-core potential
associated with attractive and repulsive dipolar interaction can be conveniently parameterized by the corresponding average characteristic lengths $\bar{\lambda}_{\pm}$, which, consistently with the definition of the effective excluded volume in Eq. \eqref{Vex_op} and following Eqs. \eqref{etabar}-\eqref{etabar1}, are defined by
\begin{equation}
	\left(\frac{\bar{\lambda}_{\pm}}{l_0}\right)^d
	=
	\int_{I_{\pm}}
	d\Delta_{ij}\,
	P(\Delta_{ij}|I_{\pm})\,
	e^{-\bar{\chi}_0(\rho)\Delta_{ij}},
	\label{lambdabar}
\end{equation}
where $P(\Delta_{ij}|I_{\pm})$ is the probability distribution introduced
in Eq. \eqref{PDeltaI}, and
\begin{equation}
	I_- = (-2,0),
	\qquad
	I_+ = (0,2).
\end{equation}
The resulting $\left(\frac{\bar{\lambda}_{\pm}}{l_0}\right)^d$ are shown in
Fig. \ref{fig:TwoScaleLength} for three representative thermodynamic states, two
paraelectric states with different densities and one ferroelectric
state.
It is worth noting that, according to Eq. \eqref{lambdabar}, $\bar{\lambda}_{\pm}$ depend
on the density through
$\chi_0(\rho)$, and on the dipolar configurations through $P(\Delta_{ij}|I_{\pm})$. $\bar{\lambda}_{\pm}$ is therefore sensitive to both density and the onset of dipolar order, as illustrated in
Fig. \ref{fig:TwoScaleLength}.
Upon averaging
over the two ranges of $\Delta_{ij}$, as shown in Fig. \ref{fig:TwoScaleLength} the adaptive soft-core interaction
in Eq. \eqref{v0softlambdaij} can be effectively represented by a repulsive potential with two
characteristic length scales: one associated with
$\Delta_{ij}\in I_{-}$ and the other with $\Delta_{ij}\in I_{+}$.
The resulting effective interaction thus belongs to the class of
two-length-scale repulsive core-softened potentials, including Jagla-like
potentials \cite{Jagla, Buldyrev}. These potentials are known to exhibit liquid-liquid phase transitions \cite{Malescio,Ryzhov,Ricci}, for which the presence of two repulsive length scales is thought to be a key ingredient \cite{Ryzhov,Malescio,Buldyrev}.
For the sake of simplicity, the present treatment retains only the
linear dependence of $\lambda_{ij}$ on $\Delta_{ij}$ in
Eq. \eqref{lambdaij}, neglecting the anharmonic corrections in
Eq. \eqref{lij_anh}. Including these corrections would make the dependence of $\lambda_{ij}$ on $\Delta_{ij}$ asymmetric, progressively suppressing the contraction of $\lambda_{ij}$ below $l_0$.
In the limiting case of a hard-core reference potential $v_0$, the
effective interaction would consist of a hard core plus a soft
repulsive shoulder, similar to the Jagla-like potential analysed
in Ref. \cite{Ryzhov}.
Two-scale isotropic soft-core potentials have been proposed as effective representations of short-range pair interaction potentials in systems with anisotropic interactions, such as water \cite{Sadr,Sadr1,Scala}, where directional hydrogen bonding makes the characteristic interparticle distance dependent on molecular orientation \cite{Stillinger}.

The liquid-liquid phase transition observed in the paraelectric phase,
see Sec. \ref{SecIIb}, occurs between two phases with the same isotropic
dipolar orientational distribution. The two characteristic length scales
of the effective soft-core repulsion are therefore controlled by density
through $\bar \chi_0(\rho)$, rather than by a change
in dipolar order. Specifically, $\bar \chi_0(\rho)$ reweights the different
values of $\Delta_{ij}$ through the factor
$e^{-\bar \chi_0(\rho)\Delta_{ij}}$ in the averages in Eq. \eqref{lambdabar}, thereby shifting
the two emergent length scales. This mechanism provides an analogue
of the compression-induced redistribution of populations between local configurations
associated with the two length scales in Jagla-like potentials, which underlies the liquid-liquid phase transition \cite{Franzese}. The present theory thus establishes two-scale soft-core repulsion as an emergent phenomenon in dipolar liquids, rather than a phenomenological assumption as in Jagla-like potentials.
A further original outcome of the present theory follows from the
dependence of the effective excluded volume, and thus of $\lambda_{\pm}$ on both dipolar order,
encoded in the probability distribution of $\Delta_{ij}$, and density,
through $\chi_0(\rho)$. 
This setting allows the ferroelectric phase transition to drive a liquid-liquid phase transition, as shown in Sec. \ref{SecIIa}. The two phenomena, namely a ferroelectric phase transition driving a liquid-liquid phase
transition and a liquid-liquid phase transition arising in the
paraelectric phase potentially inducing ferroelectric order, emerge from the same theoretical treatment of
dipolar liquids. The occurrence of either phenomenon depends on the interaction parameters
entering $v_0$ and $w_p$, the constitutive law of $\chi_0(\rho)$, and the
thermodynamic conditions.

\section{Resolving dipolar-interaction-dependent contact distances in supercooled water} \label{SecIII}

As shown in Sec. \ref{SecIIa},
$\lambda_{ij}$ in Eq. \eqref{lambdaij} is encoded in $g(h\mid\Delta_{ij}\in I)$, with the four intervals
$I$ defined in Eq. \eqref{intervals} partitioning the
$\Delta_{ij}<0$ and $\Delta_{ij}>0$ ranges, see Fig. \ref{fig:conditional_gr}. In particular, Sec. \ref{SecIIa} analyzes $g(h\mid\Delta_{ij}\in I)$ and $g(h)$ in a paraelectric (high-density) and a ferroelectric (low-density) state, whose densities differ by about $10\%$.
To assess whether similar signatures can be identified in the radial pair correlations of a realistic system, MD simulations of TIP4P/Ice water in the NPT ensemble were performed. Details are provided in Appendix \ref{app:MD}.
Representative states of the low-density liquid (LDL), at
$T=196$ K, $P=0$ bar, and $\rho=0.992\pm0.005~\mathrm{g\,cm^{-3}}$,
and of the high-density liquid (HDL), at $T=196$ K, $P=2500$ bar,
and $\rho=1.111\pm0.002~\mathrm{g\,cm^{-3}}$, were considered.
The two states differ in density by approximately $10\%$.
The LDL and HDL states exhibit ferroelectric and paraelectric behavior,
respectively, consistently with the analysis of
Refs. \cite{Izzo1,Malosso}. The
corresponding fourth-order polarization Binder cumulants \cite{Binder} are
$BC=0.655\pm0.008$ for the LDL and $BC=0.46\pm0.10$ for the HDL. Statistical errors are estimated by block averaging. Consistently with the treatment of dipolar liquids in Secs. \ref{SecI} and \ref{SecII}, intermolecular distances are measured between the molecular dipolar centres, each defined as the midpoint between the centres of positive and negative charge, with the corresponding intermolecular vector entering $\Delta_{ij}$. For each state, $g(r\mid\Delta_{ij}\in I)$ was evaluated together with $g(r)$. For each interval $I$, $g(r\mid\Delta_{ij}\in I)$ was obtained by averaging the $\Delta_{ij}$-resolved radial pair distribution functions over $I$, weighted by the fraction of pairs in each $\Delta_{ij}$ bin within $I$.
\begin{figure*}[t]
	\centering
	\includegraphics[width=14.6cm]{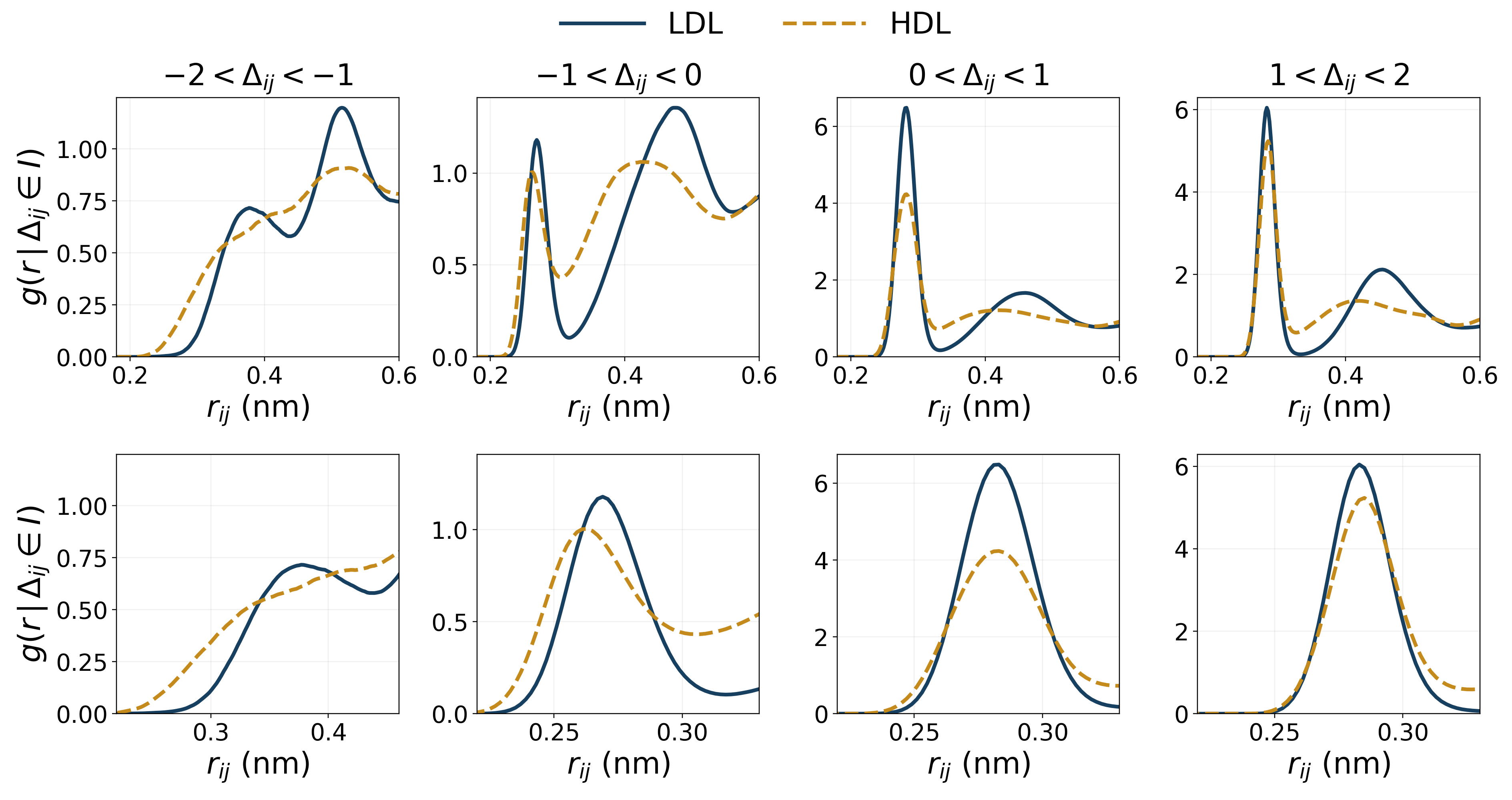}
	\caption{
		$g(r\mid\Delta_{ij}\in I)$ for the LDL and
		HDL states of TIP4P/Ice water. The four intervals $I$ are defined in Eq. \eqref{intervals}. The lower panels show a magnification of the first coordination shell.
	}
	\label{fig:conditional_gr_water}
\end{figure*} 
\begin{figure}[t]
	\centering
	\includegraphics[width=4cm]{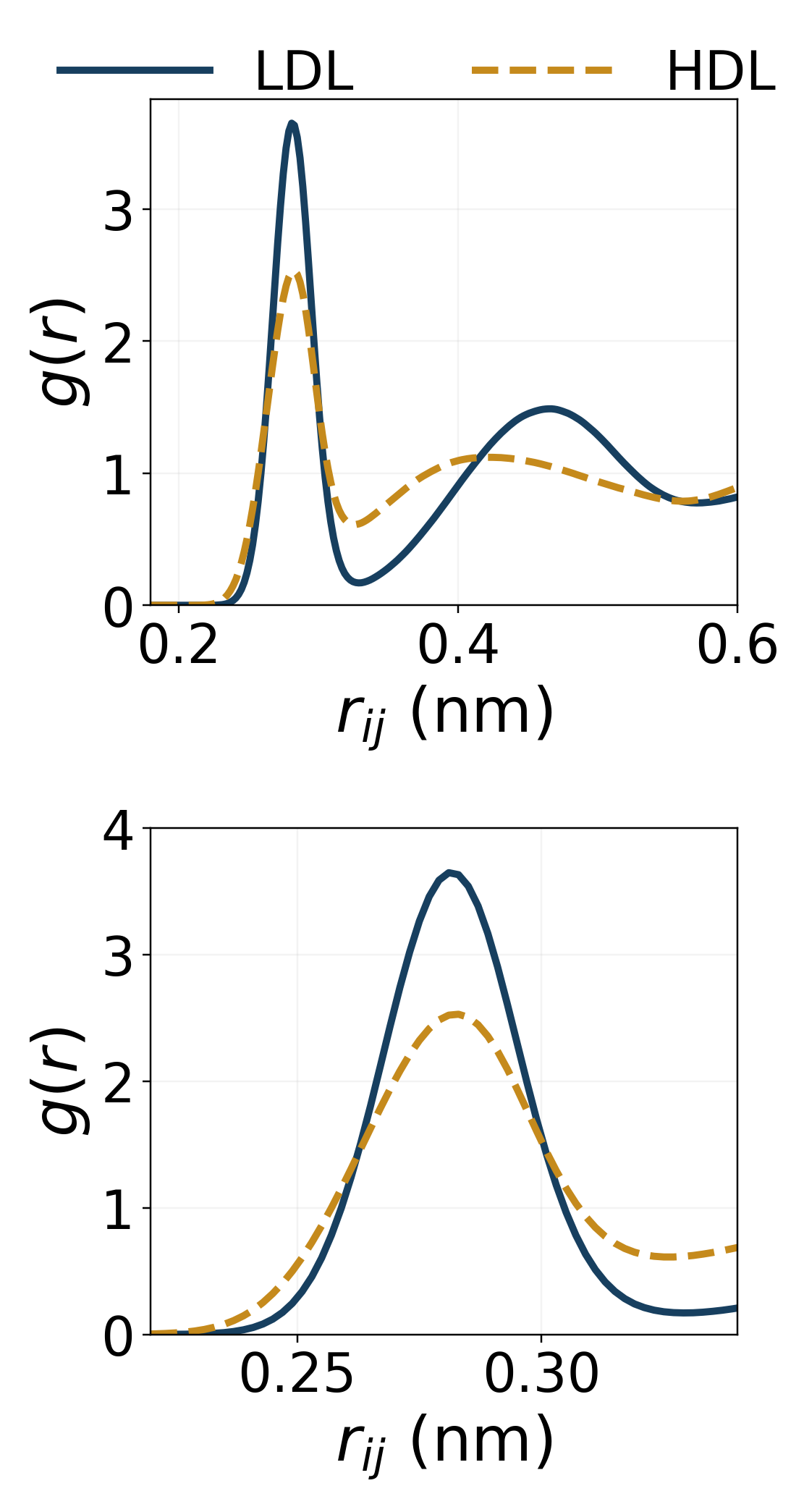}
	\caption{
		$g(r)$ for the LDL
		 and HDL  states of
		TIP4P/Ice water. The lower panel shows a magnification of the first coordination shell.
	}
	\label{fig:gr_total_water}
\end{figure}
The resulting $g(r\mid\Delta_{ij}\in I)$ are shown in Fig. \ref{fig:conditional_gr_water}. The upper panels reveal two main features: (i) the sign of $\Delta_{ij}$ discriminates between the first two coordination shells. Molecular pairs with attractive dipolar interactions ($\Delta_{ij}>0$) predominantly populate the first shell, whereas pairs with repulsive dipolar interactions ($\Delta_{ij}<0$) are preferentially found in the second shell, consistently with previous studies relating the radial distribution of neighboring water molecules to their relative orientation and pair interaction energy \cite{Svishchev,Mathias};
(ii) this overall partition is preserved in both liquid states, while the corresponding radial correlations are enhanced in the LDL, with sharper and more pronounced first- and second-shell features. This behavior is consistent with the enhanced tetrahedral ordering generally associated with the LDL state \cite{Nilsson,Gallo}.
The most striking result, however, emerges from the lower panels of
Fig. \ref{fig:conditional_gr_water}, which resolve the first coordination
shell in greater detail. 
In both the HDL and LDL states, molecular pairs with $\Delta_{ij}<0$ are found at systematically larger separations than pairs with $\Delta_{ij}>0$.
Remarkably, this difference is enhanced in the LDL, with $\Delta_{ij}<0$ and $\Delta_{ij}>0$ correlations shifting toward larger and shorter distances relative to HDL, respectively.
These results provide qualitative support for the dipolar-interaction-induced deformation of the contact distance introduced through $\lambda_{ij}$ in Eq. \eqref{lambdaij}, as well as for its density dependence described by $\chi_0(\rho)$ through Eq. \eqref{chi0_rho_negative}. Indeed, $\chi_0>0$ increases with decreasing density, enhancing the difference between the effective contact distances associated with $\Delta_{ij}>0$ and $\Delta_{ij}<0$.
This behavior is also reflected in the $g(h\mid\Delta_{ij}\in I)$ predicted by the coarse-grained dipolar-liquid theory in Fig. \ref{fig:conditional_gr}, in qualitative agreement with the results for supercooled water shown in Fig. \ref{fig:conditional_gr_water}.
The less pronounced shift toward shorter distances for $\Delta_{ij}>0$ in Fig. \ref{fig:conditional_gr_water} compared with Fig. \ref{fig:conditional_gr} may reflect the linear dependence of $\lambda_{ij}$ on $\Delta_{ij}$ assumed in the theory. 
Higher-order terms, arising from the anharmonicity of the underlying interaction in Eq. \eqref{lij_anh}, break the symmetry between contraction and expansion of the effective contact distance around $l_0$. The steep short-range repulsion preferentially suppresses contraction.
The total $g(r)$ for HDL and LDL water obtained from MD simulations, shown in Fig. \ref{fig:gr_total_water}, exhibits no significant shift of the first peak between the two phases \cite{Foffi,Gallo}. This behavior qualitatively agrees with the compensation mechanism predicted by the coarse-grained theory, as shown in Fig. \ref{Fig2_bis_grtot}.
In water, this compensation is further reinforced by the redistribution of molecular configurations between the first and second coordination shells. In particular, the first shell is predominantly populated by pairs with $\Delta_{ij}>0$, for which the first-peak shift is smaller, as observed above.
While the density increase from LDL to HDL water is commonly associated with the progressive occupation of interstitial regions between the first and second coordination shells, with only a weak displacement of the nearest-neighbor peak in $g(r)$ \cite{Foffi, Gallo}, the present analysis points to a complementary mechanism contributing to the lower density of the LDL.
The increase in effective excluded volume is not associated with an isotropic expansion of the local contact distance, which would produce a corresponding shift of the first peak of the total $g(r)$. Rather, these observations are consistent with an enhanced anisotropy of the excluded volume. 
This anisotropy, induced by the dipolar interaction through the $\Delta_{ij}$-dependent equilibrium contact distance $\lambda_{ij}$, leaves a clear signature in $g(r\mid\Delta_{ij}\in I)$ while remaining hidden in the total $g(r)$. 
Notice that, within the present theory, a decrease in density drives $\bar\chi_0>0$ toward more positive values, thereby enhancing the anisotropy associated with $\lambda_{ij}$ while simultaneously increasing the effective excluded volume, as shown in Eq. \eqref{Vex_chi_monotonic} of Sec. \ref{SecIC}.
This mechanism is not mutually exclusive with the reorganization of the second coordination shell across the LDL-to-HDL transition. Indeed, the upper panels of Fig. \ref{fig:conditional_gr_water} shows that this rearrangement also depends on $\Delta_{ij}$, suggesting that the influence of dipolar interactions on local structure may extend beyond nearest neighbors.
The present coarse-grained theory is formulated at the two-body level: the dipolar interaction renormalizes the pair contact distance as a function of $\Delta_{ij}$.
Appendix \ref{App:tetrahedral} shows how several neighbours particles can simultaneously accommodate the preferred $\Delta_{ij}$-dependent pair contact distances and how local tetrahedral order can coexist with ferroelectric order in the liquid phase, in agreement with the observations of Ref. \cite{Yuan}. The pairwise mechanism underlying the coarse-grained theory, and the arguments developed from it in Appendix \ref{App:tetrahedral}, neglect higher-order molecular multipoles of water, hydrogen bonding and quantum effects. The aim is therefore not to explain the microscopic origin of local tetrahedral order in LDL water, but to show that such order is compatible with the elementary mechanism proposed here and can coexist with aligned dipoles and the resulting macroscopic polarization.

\begin{figure}[t]
	\centering
	\includegraphics[width=5.7cm]{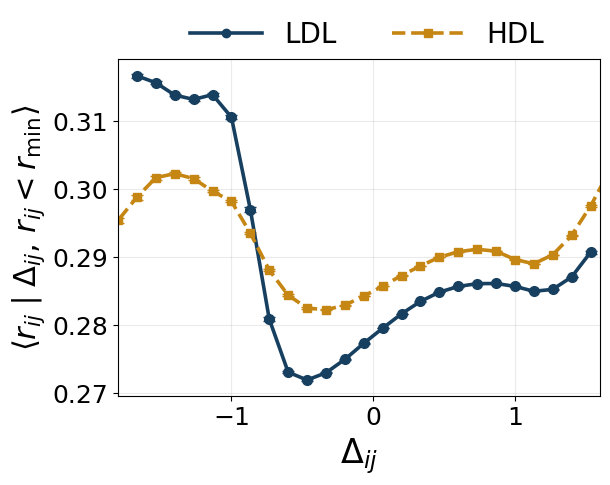}
	\caption{
		Mean intermolecular distance within the first
		coordination shell,
		$\langle r_{ij}\mid\Delta_{ij},\,r_{ij}<r_{\min}\rangle$ (
		$r_{\min}=0.33$ nm)
		as a function of $\Delta_{ij}$ for the
		LDL and HDL states of
		TIP4P/Ice water. Each point represents an average over molecular pairs in a $\Delta_{ij}$ bin of width $0.133$. Bins containing fewer than 2000 pairs are omitted. For each $\Delta_{ij}$ bin, errors bars show the standard deviation of the mean pair distances calculated separately for each contributing MD frame, divided the square root of the number of contributing frames. 
	}
	\label{fig:conditional_distance}
\end{figure}

For a further characterization of the behaviour discussed above,
Fig. \ref{fig:conditional_distance} shows the mean intermolecular distance
within the first coordination shell,
$\langle r_{ij}\mid\Delta_{ij},\,r_{ij}<r_{\min}\rangle$.
For strongly negative $\Delta_{ij}$, the mean first-neighbor distance is larger in the
LDL than in the HDL, whereas the opposite trend is observed for
$\Delta_{ij}>0$. This behavior is consistent with the corresponding shifts of the first-coordination-shell peak in $g(r\mid\Delta_{ij} \in I)$. The nonlinear dependence on $\Delta_{ij}$ observed in Fig. \ref{fig:conditional_distance} is consistent with nonlinear corrections to $\lambda_{ij}$ arising from the anharmonicity of the interaction potentials, as in Eq. \eqref{lij_anh}, which may also account for the weaker shift toward shorter distances for $\Delta_{ij}>0$ discussed above. The minimum, however, occurs at negative $\Delta_{ij}$ rather than at $\Delta_{ij}=0$, a feature not captured by the present coarse-grained theory. It may reflect many-body effects on the local molecular arrangement. 

The Supplemental Note II reports the corresponding results for two
additional LDL-HDL pair. The first consists of two states at $P=0$ bar and different $T$. The second consists of the reference states analyzed in Ref. \cite{Izzo1}, drawn from
the MD simulations of Ref. \cite{Debenedetti}.

These observations provide evidence that the elementary mechanism captured by the coarse-grained theory may be operative in supercooled water. In particular, the TIP4P/Ice simulations reveal distinct LDL and HDL responses to the local dipolar arrangement, consistent with the coarse-grained theory. However, atomistic MD simulations do not allow this mechanism to be isolated from the other effects at play in water and therefore cannot establish a causal link between the microscopic mechanism underlying the coarse-grained theory and the occurrence of the liquid-liquid phase transition in supercooled water. Such a causal link is established only within the coarse-grained theory as solved in the $d \rightarrow \infty$ limit. 
Two complementary routes could be pursued to investigate the proposed mechanism in supercooled water. First, $\chi_0(\rho)$ could be inferred from ensembles of liquid inherent structures at different densities. Second, machine-learning-assisted coarse graining of the atomistic interaction potential \cite{Wang,Majewski} could be performed using the functional form proposed here, possibly including the leading nonlinear contribution in $\Delta_{ij}$. Coarse-grained MD simulations based on either parametrization would then test whether the proposed mechanism is sufficient to reproduce the liquid-liquid phase transition of supercooled water.

\section{Conclusions} \label{Concl}
Building on the idea that directional interactions can shape the local structure of a liquid, as shown for water through the analysis of its  inherent structures \cite{Stillinger}, a coarse-grained theory of dipolar liquids is developed in which the characteristic length of the effective short-range repulsive interaction, which sets the contact distance between a pair of particles and hence their excluded volume, depends on the angular factor $\Delta_{ij}$ of the dipolar interaction.
The underlying mechanism bears an analogy to ferroelastic coupling in crystals \cite{Chaikin}. 
The effective pair contact distance $\lambda_{ij}$ thus provides a coarse-grained structural coordinate that parametrizes a state-dependent soft-core effective pair interaction, which is assumed to incorporate the many-body response of the translational degrees of freedom. The dipolar interaction, however, is retained explicitly and, at finite dimension, generates higher-order virial contributions to the free energy. The large-dimensional limit provides a convenient setting for assessing the consequences of this coarse-grained description, since these higher-order virial contributions vanish and the excess free energy becomes exactly determined by the second-virial term.
Notably, the dipolar interaction is anisotropic in the space of the translational coordinates. Making the pair contact distance dependent on the dipolar interaction therefore translates this anisotropy into an anisotropic effective excluded volume.
Within the present coarse-grained theory, the coupling between the equilibrium pair contact distance and $\Delta_{ij}$ is controlled by the structural compliance $\chi_0$, which is inversely proportional to the effective density-dependent stiffness $k_2^{\rm eff}(\rho)$. Local stability then implies $\chi_0>0$, while the physically motivated assumption of compression-induced stiffening, $d k_2^{\rm eff}/d\rho>0$, gives $d\chi_0/d\rho<0$. A more general coarse-grained description, in which many-body effects also renormalize the coupling between $\lambda_{ij}$ and $\Delta_{ij}$, can also allow $\chi_0<0$, while compression-induced softening can give $d\chi_0/d\rho>0$. 
The signs of $\chi_0$ and its density derivative affect the predictions of the theory in distinct ways. They can qualitatively modify the predicted ferroelectric phase transition, possibly favouring antiferroelectric ordering, whereas their effect on the liquid-liquid phase transition within the paraelectric phase is only quantitative, since the mechanism allowing this transition relies on the density dependence of $\chi_0$. The trends observed in the pair correlation functions conditioned on $\Delta_{ij}$ in the TIP4P/Ice MD simulations are, however, consistent with $\chi_0>0$ and $d\chi_0/d\rho<0$. The dependence of $\lambda_{ij}$ on $\Delta_{ij}$ is assumed to be linear for the sake of simplicity. Anharmonicity of realistic interaction potentials, such as the Lennard-Jones potential, generally gives rise to nonlinear contributions to this dependence. As shown in Supplementary Note I, inclusion of the leading nonlinear contribution does not qualitatively modify the predictions of the theory for the phase transitions. The nonlinear dependence of $\langle r_{ij}\mid\Delta_{ij},\,r_{ij}<r_{\min}\rangle$ observed in TIP4P/Ice MD simulations, as shown in Fig. \ref{fig:conditional_distance}, is consistent with the occurrence of such corrections.

Variational minimization of the coarse-grained free energy for small polarization and structural compliance reveals the onset of a bulk ferroelectric phase transition. 
The dipolar interaction is unchanged from Ref. \cite{Izzo} and is unaffected by the coarse-grained representation. 
The ferroelectric phase transition also drives a liquid-liquid phase transition between paraelectric and ferroelectric phases of distinct densities.  
A closely related result was obtained in Ref. \cite{Izzo1}, where modelling supercooled water as a dipolar liquid led to the conclusion that a ferroelectric phase transition can underlie the liquid-liquid phase transition. There, this interplay was traced at the macroscopic level to the density dependence of the dipolar-interaction contribution to the excess free energy within a classical DFT framework.
The present coarse-grained theory identifies a possible microscopic mechanism for this interplay, linking the onset of ferroelectric ordering to a local structural reorganization and the resulting variations in the effective excluded volume and the density.
The present theory further generalizes this scenario by showing that the sign and magnitude of the density change depend on the critical density and temperature of the ferroelectric phase  transition, as well as on the interaction-potential parameters and the parameters defining $\chi_0$, allowing dipolar ordering to produce either an increase or a decrease in density. For the parameter values considered in the present study, the ferroelectric phase becomes less dense at higher critical densities. The high-density, low-temperature regime relevant to supercooled water is therefore consistent with the occurrence of a less dense ferroelectric phase within the present theory. The possibility of a tricritical point and a first-order ferroelectric liquid-liquid phase transition, discussed in Ref. \cite{Izzo1}, lies beyond the present analysis. The occurrence of a ferroelectric phase transition is linked to the sign of $\chi_0$. While $\chi_0>0$ allows a ferroelectric phase transition, small negative values of $\chi_0$ may instead lead to an antiferroelectric phase transition. For $\chi_0>0$, a positive density derivative, $d\chi_0/d\rho>0$, implies that the liquid-liquid phase transition accompanying the ferroelectric phase transition is always characterized by an increase in the equilibrium density upon ferroelectric ordering.

The coarse-grained theory further predicts a first-order liquid-liquid phase transition within the paraelectric phase, whose occurrence relies on the density dependence of $\chi_0$. Its quantitative properties are governed by the interaction-potential parameters and how $\chi_0$ varies with density. For dipolar liquids, the Jagla-like two-length-scale core-softened potentials, known to support liquid-liquid phase transitions \cite{Jagla,Franzese,Ricci}, can be derived as effective representations of the anisotropic soft-core interaction underlying the coarse-grained theory, as discussed in Sec. \ref{SecII3}. This provides a rationale for the link between directional interactions and two-length-scale effective potentials previously hypothesized for water \cite{Stillinger,Sadr,Sadr1,Scala}. Under appropriate thermodynamic conditions, the liquid-liquid phase transition within the paraelectric phase can in turn promote the onset of ferroelectric order.
As shown in Fig. \ref{fig:critical_density}, the first-order liquid-liquid coexistence line has a positive slope in the pressure-temperature plane. According to the Clapeyron relation, the sign of this slope is determined by the entropy and volume differences between the coexisting liquid phases, $\Delta S$ and $\Delta V$ respectively,   \cite{Ricci}, $\left(dP/dT\right)_{\rm coex}=\Delta S/\Delta V$.  The slope is therefore not universal and depends on the microscopic interactions and thermodynamic conditions. In the case of supercooled water, the liquid-liquid coexistence line has been found to have a negative slope \cite{Palmer}. 
Interestingly, if the first-order liquid-liquid transition predicted by the present theory were accompanied by a ferroelectric phase transition, the entropy reduction associated with ferroelectric ordering could change the sign of $\Delta S$ and reverse the slope of the coexistence line.
The same coarse-grained theory thus predicts two distinct scenarios for dipolar liquids: a ferroelectric phase transition driving a liquid-liquid phase transition, and a first-order liquid-liquid phase transition emerging within the paraelectric phase and potentially promoting ferroelectric order. Which scenario occurs depends on the potential-interaction parameters, the density dependence of the structural compliance, and the thermodynamic conditions.
This second scenario may provide an alternative picture of the phase behaviour of supercooled water to that in Ref. \cite{Izzo1}. Distinguishing between these scenarios requires identifying the order parameter of the liquid-liquid phase transition in supercooled water, through a finite-size analysis of the polarization and its Binder cumulants, density and related susceptibilities, across the thermodynamic plane.

It would be of interest to test whether the phase-transition predictions obtained in the $d\to\infty$ limit remain valid in finite dimensions, through numerical simulations of the coarse-grained model in three dimensions with appropriately selected parameters.

Within the present theory, the dependence of $\lambda_{ij}$ on $\Delta_{ij}$ leaves a clear signature in radial pair correlation functions conditioned on $\Delta_{ij}$. Interestingly, this signature is largely hidden in the total radial pair correlation function, where the anisotropy of the effective excluded volume is averaged out. Consequently, changes in effective excluded volume between low- and high-density liquid phases that are associated with changes in its anisotropy are more clearly reflected in the conditional than in the total pair correlation functions. Remarkably, this behavior is also found in MD simulations of LDL and HDL states of TIP4P/Ice water.
This observation reveals that the liquid-liquid phase transition between HDL and LDL in supercooled water involves not only the reorganization of the interstitial region between the first and second coordination shells \cite{Foffi,Gallo,Nilsson}, but also structural changes within the first coordination shell itself. In particular, the conditional pair correlations point to an enhanced spatial anisotropy of the first coordination shell in the LDL, associated within the present coarse-grained theory with an increase in effective excluded volume. These findings thus support a possible role of dipolar interaction in shaping the effective excluded volume of supercooled water and its evolution between the two liquid phases. Appendix \ref{App:tetrahedral} shows that the pair-level mechanism underlying the coarse-grained theory can generate a many-particle arrangement in which local tetrahedral structure coexists with ferroelectric order in the liquid phase. Locally tetrahedral structures exhibiting local polarization have recently been identified in MD simulations of interfacial supercooled TIP4P/2005 water \cite{Yuan}. The TIP4P/Ice MD simulations provide evidence that the elementary mechanism underlying the coarse-grained theory may be operative in supercooled water. These atomistic MD simulations, however, cannot isolate this mechanism from the other effects at play and therefore cannot establish a causal link between the mechanism and the liquid-liquid phase transition observed in supercooled water. Coarse-grained MD simulations parametrized from atomistic water could test whether the proposed mechanism is sufficient to reproduce the liquid-liquid phase transition of supercooled water.

\section*{Supplementary Material}
Supplementary Note I extends the coarse-grained theory by including the leading nonlinear, anharmonic correction in $\Delta_{ij}$ to the effective soft-core length scale $\lambda_{ij}$. It shows that the ferroelectric phase transition, its accompanying liquid-liquid phase transition, and the liquid-liquid phase transition within the paraelectric phase can persist, while their quantitative features are modified.
Supplementary Note II reports $g(r|\Delta_{ij}\in I)$ in the four intervals in Eq. \eqref{intervals}, $g(r)$, and $\langle r_{ij}\mid\Delta_{ij},\,r_{ij}<r_{\min}\rangle$ obtained from MD simulations of TIP4P/Ice water for an LDL-HDL pair along the $P=0$ bar isobar, at $T=196$ K and $T=270$ K, and for the pair of representative LDL and HDL states analyzed in Ref. \cite{Izzo1}, based on simulation data from Ref. \cite{Debenedetti}.

\begin{acknowledgments}
	Support from Ministero Istruzione Universitá Ricerca - Progetti di Rilevante Interesse Nazionale “Deeping our understanding of the Liquid–Liquid transition in supercooled water”, grant 2022JWAF7Y, is acknowledged. Computational resources were provided by CINECA under the ISCRA-C programme (projects IsCd9\_FLLPTW and IsCb8\_LLPTW).
\end{acknowledgments}

\appendix

\section{Numerical evaluation of the paraelectric and ferroelectric reference states for the model pair correlation functions}
\label{AppendixI}

For a chosen value of the paraelectric reduced density $\eta_P=16$ the
critical inverse temperature $\beta_c$ is determined numerically by
solving
\begin{equation}
	A(\eta_P,\beta_c)=0,
	\label{eq:numerical_critical_condition}
\end{equation}
where $A(\eta,\beta)$ is the coefficient of the $\delta^2$ term in the
Landau expansion of the free energy, defined in
Eq. \eqref{A_rho_beta}.
The corresponding critical pressure is then obtained from the full
isobaric equation of state, Eq. \eqref{EOS_chi_rho},
\begin{equation}
	P_c
	=
	P(\eta_P,\beta_c,\delta=0).
	\label{eq:P_critical_numerical}
\end{equation}
The structural compliance entering the equation of state is evaluated at
the paraelectric density according to the constitutive relation,
Eq. \eqref{chi0exp},
\begin{equation}
	\bar{\chi}_0^P
	=
	\bar{\chi}_0(\eta_P).
\end{equation}
The corresponding reduced density $\eta_F$ of the ferroelectric reference state at $\delta=1$ is obtained by
solving
\begin{equation}
	P(\eta_F,\beta_c,\delta=1)
	=
	P_c,
	\label{eq:isobaric_condition}
\end{equation}
where $P_c$ is the paraelectric critical pressure defined in
Eq. \eqref{eq:P_critical_numerical}.
Among the solutions of
Eq. \eqref{eq:isobaric_condition}, only mechanically stable states
satisfying
\begin{equation}
	\left.
	\frac{\partial P}{\partial\eta}
	\right|_{\eta_F,\beta_c,\delta=1}
	>0
	\label{eq:mechanical_stability}
\end{equation}
are retained.
The numerical solution finally provides the ferroelectric reduced
density $\eta_F$, the corresponding structural compliance
\begin{equation}
	\bar{\chi}_0^F
	=
	\bar{\chi}_0(\eta_F),
\end{equation}
and the relative density variation
\begin{equation}
	\frac{\Delta\rho}{\rho_P}
	=
	\frac{\eta_F-\eta_P}{\eta_P}.
\end{equation}

\section{MD simulations}
\label{app:MD}

MD simulations were performed with the GROMACS 2023 package for
$N=1000$ TIP4P/Ice water molecules in the NPT 
ensemble using a Nos\'e-Hoover thermostat
($\tau_T=8.011$ ps) and an isotropic Parrinello-Rahman barostat
($\tau_P=18.3$ ps). The isothermal compressibility was
set to $4.5\times10^{-5}$ bar$^{-1}$. The integration time step was
$2$ fs, with all bonds constrained using the LINCS algorithm.
Short-range van der Waals and electrostatic interactions were truncated
at $0.9$ nm, while long-range electrostatics were treated using the
particle-mesh Ewald (PME) method with a Fourier spacing of $0.1$ nm and
a relative Ewald tolerance of $10^{-5}$. Long-range dispersion corrections were applied to both energy and pressure. The MD trajectory was $4\,\mu$s long.  Pair correlation functions
were computed from configurations sampled every $0.8$ ns along the MD trajectory.

\section{Ferroelectric Ordering and Local Tetrahedral Structure}
\label{App:tetrahedral}
In the coarse-grained description introduced in the main text, the
effective length $\lambda_{ij}$ provides a pair-level description of
the local structural response of a macroscopically homogeneous liquid.
While this construction determines the preferred local arrangement of
an isolated pair, it does not determine how these preferred pair arrangements can be simultaneously realized by multiple particles to generate the local structure of the liquid. In this Appendix, this issue is examined through a simple three-dimensional geometrical construction.

\begin{figure}[t]
	\centering
	\includegraphics[width=0.75\columnwidth]{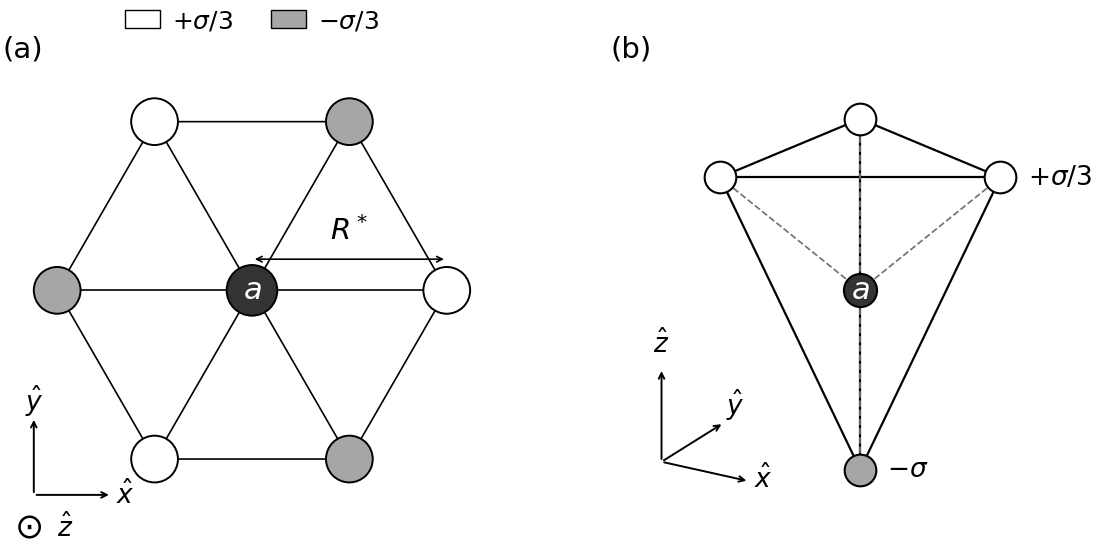}
	\caption{(a) Local arrangement of parallel chains in the plane
		perpendicular to the polarization direction $\hat z$. The central chain $a$ is surrounded by six neighboring chains at transverse distance $R^*$. White and gray circles denote axial shifts $+\sigma/3$ and $-\sigma/3$, respectively, relative to chain $a$. (b) One of the two tetrahedra formed by contact neighbors around a reference particle on chain $a$. Three neighbors lie at axial displacement
		$+\sigma/3$, while the fourth is the nearest neighbor on the same
		chain at $-\sigma$. Solid lines connect the four vertices of the
		tetrahedron, while dashed lines connect the reference particle to
		the four neighbors}
	\label{fig:chain_local}
\end{figure}

The dipolar interaction favors head-to-tail chains independently of occurrence of paraelectric or ferroelectric phase \cite{Klapp}, as encoded by $\lambda_{ij}$ for $\chi_0>0$. The local structure of the ferroelectric phase is examined in the
fully polarized limit, with the polarization chosen along the $z$
direction,
\begin{equation}
	\hat d_i=\hat z
	\qquad \forall i .
\end{equation}
In this limit, the head-to-tail chains are oriented along the
polarization direction and can therefore be represented as parallel
chains along $z$.
The particles are assigned a hard-core diameter $\sigma$, providing a
lower bound to the effective soft-core length $\lambda_{ij}$ introduced
in the main text. The dipolar-induced contraction of $\lambda_{ij}$ is
then limited by $\sigma$, which fixes the nearest-neighbor separation
between particles belonging to the same chain at maximal packing.
As a minimal construction, consider two chains, $a$ and $b$, oriented
along $\hat z$.
Their axes are separated by a transverse distance $R_{ab}$, while the
particles of one chain are displaced by a distance $s_{ab}$ along
$\hat z$ relative to those of the other chain. By periodicity and
symmetry, inequivalent relative shifts satisfy
$0\leq s_{ab}\leq\sigma/2$. For a given $s_{ab}$, the hard-core
condition requires
\begin{equation}
	R_{ab}^2+s_{ab}^2\geq\sigma^2.
\end{equation}
Restricting to configurations at maximal packing,
\begin{equation}
	R_{ab}(s_{ab})=\sqrt{\sigma^2-s_{ab}^2}.
\end{equation}
Since $R_{ab}(s_{ab})$ decreases over the allowed range of $s_{ab}$,
maximal packing is obtained for
\begin{equation}
	s_{ab}^\star=\frac{\sigma}{2},
	\qquad
	R_{ab}^\star=\frac{\sqrt{3}}{2}\sigma.
\end{equation}
Consider then three parallel chains, $a$, $b$, and $c$, whose axes form
an equilateral triangle in the plane perpendicular to the polarization
direction, so that
\begin{equation}
	R_{ab}=R_{bc}=R_{ac} \equiv R.
\end{equation}
This construction is not intended to specify a distinct thermodynamic phase, but represents a local motif that can occur in the dense liquid.
If each pair were considered independently, maximal packing would
require a relative shift $\sigma/2$. Taking chain $a$ as reference,
the pairwise optimal shifts $s_{ab}=s_{bc}=\sigma/2$ imply
\begin{equation}
	s_{ac}=s_{ab}+s_{bc}=\sigma. \label{shift3chains}
\end{equation}
Since the particles along each chain are separated by $\sigma$, chains
$a$ and $c$ are therefore in phase and cannot simultaneously realize
the optimal two-chain shift. The packing of three neighboring chains
is thus geometrically frustrated. The maximal-packing configuration is then determined by optimizing $s_{ij}$, $ij=ab,bc,ac$ and $R$. Since the particle positions are periodic with period $\sigma$, the
corresponding inequivalent shift is
\begin{equation}
	s_{ac}=\min\left\{
	s_{ab}+s_{bc},
	\,\sigma-s_{ab}-s_{bc}
	\right\}. \label{shift_constrain}
\end{equation}
The hard-core condition requires
\begin{equation}
	R \geq \sqrt{\sigma^2-s_{ij}^2},
	\qquad ij=ab,bc,ac.
\end{equation}
Since $\sqrt{\sigma^2-s_{ij}^2}$ decreases with increasing $s_{ij}$,
the chain pair with the smallest $s_{ij}$ sets the minimum allowed
value of $R$. Maximal packing therefore
requires maximizing
\begin{equation}
	\min\{s_{ab},s_{bc},s_{ac}\}.
\end{equation}
From Eq. \ref{shift_constrain}, the optimum is obtained when
\begin{equation}
	s_{ab}=s_{bc}=s_{ac}\equiv s^*=\frac{\sigma}{3},
\end{equation}
yielding
\begin{equation}
	R^*=\frac{2\sqrt{2}}{3}\sigma.
\end{equation}
Indeed, by Eq. \eqref{shift_constrain} if both $s_{ab}$ and $s_{bc}$ exceed $\sigma/3$, then $s_{ac}<\sigma/3$, whereas setting $s_{ab}=s_{bc}=\sigma/3$ gives $s_{ac}=\sigma/3$. The three-chain arrangement can be extended to the local environment
of a given chain, as illustrated in
Fig. \ref{fig:chain_local}. A central chain $a$ can be surrounded by
six neighboring chains with alternating axial shifts $+\sigma/3$ and
$-\sigma/3$. Each triangle formed by $a$ and two adjacent neighboring chains realizes the three-chain configuration derived above, with the same $R^*$ and $s^*=\sigma/3$.
This local arrangement contains an exact tetrahedral geometry. Consider a reference particle on chain $a$ and its nearest neighbor
along the same chain at $-\sigma\hat z$. Among the six neighboring
chains, the three shifted by $+\sigma/3$ provide three additional
contact neighbors. Their projections onto the transverse plane lie at
the distance $R^*$ from chain $a$ and have azimuthal angles
$\phi_k=0,2\pi/3,4\pi/3$. Since $R^*=2\sqrt{2}\sigma/3$, each of the three neighbors on the
surrounding chains lies at a distance
\begin{equation}
	\sqrt{(R^*)^2+\left(\frac{\sigma}{3}\right)^2}=\sigma
\end{equation}
from the reference particle, as does its nearest neighbor on chain $a$
at $-\sigma\hat z$. The four neighbors are therefore equidistant from the
reference particle. The unit vectors from the reference particle to the four neighbors are
\begin{equation}
	\mathbf n_0=(0,0,-1),\qquad
	\mathbf n_k=
	\Big(
	\frac{2\sqrt{2}}{3}\cos\phi_k,
	\frac{2\sqrt{2}}{3}\sin\phi_k,
	\frac{1}{3}
	\Big).
\end{equation}
Their pairwise scalar products are
\begin{equation}
	\mathbf n_i\cdot\mathbf n_j=-\frac{1}{3},
	\qquad i\neq j,
\end{equation}
corresponding to the tetrahedral angle. Hence, the four neighbors form a regular tetrahedron
centered on the reference particle.

\end{document}